\documentclass[10pt,aps,twocolumn,prd,superscriptaddress,showpacs,nofootinbib,noshowkeys,floatfix,preprintnumbers]{revtex4}
\usepackage{times}

\usepackage[utf8]{inputenc}

\usepackage{graphics,graphicx}
\usepackage{float}
\usepackage{amsmath, amssymb}
\usepackage{multirow}
\usepackage{longtable}
\usepackage{color}
\usepackage{slashed}

\usepackage[colorlinks=true,linktocpage=true,linkcolor=blue,citecolor=blue,hyperfootnotes=true]{hyperref}
\usepackage[normalem]{ulem}  
\usepackage{subfigure}

\begin{document}


\title{Quark-Nuclear Hybrid Star EoS with Excluded Volume Effects}

\author{Mark Alexander Randolph Kaltenborn}
	\affiliation{Department of Physics, The George Washington University, Washington DC 20052, USA}
	\affiliation{Los Alamos National Laboratory, Los Alamos, NM 87545, USA}
\author{Niels-Uwe Friedrich Bastian}
	\affiliation{Institute of Theoretical Physics, University of Wroc{\l}aw, 50-204 Wroc{\l}aw, Poland}
\author{David Bernhard Blaschke}
	\affiliation{Institute of Theoretical Physics, University of Wroc{\l}aw, 50-204 Wroc{\l}aw, Poland}
	\affiliation{Bogoliubov Laboratory for Theoretical Physics, JINR Dubna, 141980 Dubna, Russia}
	\affiliation{National Research Nuclear University (MEPhI), 115409 Moscow, Russia}

\date{\today}

\begin{abstract}
A two-phase description of the quark-nuclear matter hybrid equation of state that takes into account the effect of excluded volume in both the hadronic and the quark-matter phases is introduced.
The nuclear phase manifests a reduction of the available volume as density increases, leading to a stiffening of the matter.
The quark-matter phase displays a reduction of the effective string-tension in the confining density-functional from available volume contributions.
The nuclear equation of state is based upon the relativistic density functional model DD2 with excluded volume.
The quark-matter EoS is based upon a quasiparticle model derived from a relativistic density-functional approach and will be discussed in greater detail.
The interactions are decomposed into mean scalar and vector components.
The scalar interaction is motivated by a string potential between quarks, whereas the vector interaction potential is motivated by higher-order interactions of quarks leading to an increased stiffening at high densities.
As an application, we consider matter under compact star constraints of electric neutrality and $\beta$-equilibrium.
We obtain mass-radius relations for hybrid stars that form a third family, disconnected from the purely hadronic star branch, and fulfill the $2M_\odot$ constraint.
\end{abstract}


\maketitle

\section{Introduction}
\label{intro}

Theoretical investigations into compact star matter have seen a boom over the past decade, spurred forward by the plethora of observational data collected for the masses and radii of pulsars. The mass-radius (M-R) data has been particularly useful for constraining the dense matter equation of state (EoS) at zero temperature. For a recent review see, e.g., Ref.~\cite{Lattimer:2015nhk}.
These constraints provide an experimental sandbox for probing the properties of matter that is inaccessible in terrestrial experiments. The question investigated in this work is whether or not quark matter is manifest in the interior of compact stars by a phase transition from hadronic matter to quark matter and, if quark matter is indeed manifest, what would the implications be on the observables, as well as the QCD phase diagram.

In order to answer this question, we employ an effective, relativistic density-functional approach to both the hadronic and the quark-matter equation of state (EoS) at zero temperature.
A method for implementing quark confinement in a chiral quark-matter model is based on reviving the idea of the string-flip model (SFM) \cite{Horowitz:1985tx,Ropke:1986qs}.
According to the SFM, the confining interactions between colored constituents (modeled, e.g., through a Cornell-type confining potential) get saturated within nearest neighbors.
In hadronic matter these quark neighbors are the ones bound in a hadron, while at high densities in a multi-quark system one has to draw string connections between quarks so as to neutralize the color while keeping the total length of all strings at a minimum.
This task can be solved by simulation \cite{Horowitz:1991fn} or by effective quantum statistical methods \cite{Ropke:1986qs,Horowitz:1991ux}.
One can obtain the string length distribution function for a given density and fold a given interaction potential model with this distribution in order to obtain the correlation energy in a meanfield approximation.
Since the average string length between quarks in uniform matter is related to the scalar number density $n_\mathrm s$ by being proportional to ${n_\mathrm s^{-1/3}}$, one obtains a corresponding contribution to the energy density functional of quark matter, see also \cite{Khvorostukin:2006aw}.
Adopting the confining interaction as a scalar potential, this density functional describes the medium modification of the quark mass, being a Dirac scalar. At low densities the quark mass diverges thus mimicking confinement by removing quarks as dynamical degrees of freedom.
At low density the quarks are clustered into hadrons, and this quark cluster phase is energetically preferable over the uniform quark matter with diverging scalar confining interaction.
Increasing the density, quark exchange effects between hadrons (Pauli blocking) lead to strong repulsion (that can be described by an excluded hadron volume model) and to delocalization of the quark wave function, driving the system into the homogeneous quark-matter phase.
In the phase transition region, the reduction of the available volume (in which the nonperturbative dual superconductor vacuum of QCD persists which expells the color field lines by the dual Meissner effect) leads to a reduction of the density of color field lines and thus to a ``screening'' of the string tension by the available volume fraction factor.
At still higher densities higher order quark interactions adopted according to \cite{Benic:2014iaa,Benic:2014jia} lead to a stiffening of quark matter that allows for a stable branch of hybrid stars with quark-matter core.

In order to obtain the EoS for hybrid star matter we have developed a relativistic, self-consistent
mean-field approach within the path integral approach, based on an ansatz for the potential energy density
functional from which density dependent self-energies follow as derivatives.
The resulting quasiparticle model necessarily maintains thermodynamic consistency, since all approximation steps are done on the level of the partition function from which the expressions for thermodynamical functions
follow by derivation.


When considering the possibility of a phase transition from hadronic matter to quark matter
in compact star interiors, the phenomenon of high-mass twins presents itself.
The term mass twins refers to the existence of two compact star configurations that have the same mass
but distinctly different radii.
It is related to the occurrence of a ``third family" of compact stars, besides white dwarfs and ordinary neutron stars, which is separated from the latter by an unstable branch in the mass-radius diagram.
As discussed already by Gerlach in 1968 \cite{Gerlach:1968zz}, the third family is related to the behavior 
of the high-density EoS, which may exhibit a phase transition.
The transition to quark matter or to pion condensed matter has been considered in the context of a third family of stars by K\"ampfer  \cite{Kampfer:1981yr,Kampfer:1983we}, and later this issue was taken up again, e.g., in Refs.~\cite{Schertler:2000xq,Glendenning:1998ag, Bhattacharyya:2004fn,Agrawal:2009ad}.
When the mass twins occur 
in a mass range that covers the presently best known high mass of the pulsar PSR J0348+0432 with $M=2.01\pm 0.04~M_\odot$ \cite{Antoniadis:2013pzd}, we will speak of high-mass twin (HMT) stars.

The existence of HMTs hinges on the nature of the phase transition in these dense systems.
In order for this phenomenon to appear, the EoS of the compact star matter must have a strong first-order phase transition \cite{Blaschke:2013ana,Alford:2013aca,Benic:2014jia}
characterized by a large jump in energy density of the two phases also denoted as a large latent heat.
In order to support the mass twin phenomenon, a necessary condition is
\cite{Seidov:1971,Schaeffer:1983,Alford:2013aca}
\begin{equation}
	\frac{\Delta\varepsilon}{\varepsilon_{\rm trans}} >  \frac{1}{2} + \frac{3}{2}\frac{P_{\rm trans}}{\varepsilon_{\rm trans}}~,
\end{equation}
where $\Delta\varepsilon$ is the latent heat, and $\varepsilon_{\textrm{trans}}$  ($P_{\rm trans}$ )
is the energy density (pressure) at the onset of the deconfinement phase transition.
As mentioned, the existence of HMTs would imply the existence of at least one critical endpoint in the QCD phase diagram \cite{Blaschke:2013ana,Alvarez-Castillo:2016wqj}. The existence of high mass twins can potentially be observationally confirmed.
If the radii of two compact stars with sufficiently accurately measured high mass like, e.g.,
PSR J0348+0432 \cite{Antoniadis:2013pzd}	and PSR J1614-2230 \cite{Demorest:2010bx,Fonseca:2016tux} could be measured 
(for instance by the NICER mission, which has recently been launched and mounted on the International Space Station \cite{NICER1})
and turn out significantly different from each other, then this would present evidence for HMTs that imply a strong first-order transition in compact star matter.

In order to obtain HMTs, it is necessary to fulfill the following constraints:
\begin{itemize}
	\item a strong stiffening of nuclear matter at supersaturation densities,
	\item a sufficiently soft behavior of quark matter at the hadron-to-quark-matter phase transition,
	in order to have a sufficiently large latent heat $\Delta \varepsilon$ \cite{Benic:2014jia}, and
	\item a sufficient stiffening of quark matter above the deconfinement transition.
\end{itemize}
These properties characterize the class of hybrid EoS we are investigating in the present work.

This paper is organized in the following way.
We begin by outlining our relativistic density-functional approach in Sect.~\ref{RDF}.
In Sect.~\ref{Results}, we discuss our solutions to the Tolman-Oppenheimer-Volkoff equation with parameterizations of our hybrid EoS and how observations constrain our choices of parameterization.
A discussion and concluding remarks will be given in Sect.~\ref{Discussion} and Sect.~\ref{Conclusion}, respectively.
The appearance of a stable branch of compact stars, with quark-matter core at high masses, which is disconnected from the stable neutron star branch and is called the ``third family'' of stable hybrid stars, is a main result of the new relativistic density functional based EoS introduced in this work.

\section{Hybrid star matter with excluded volume effects}
\label{RDF}

The effective relativistic density-functional (RDF) approach has been successfully applied in describing warm, dense asymmetric nuclear matter with light clusters \cite{Typel:2009sy} as well as other systems where the fundamental interaction between constituents is not a priori known.
Thus, this RDF approach is adopted and applied here to the hybrid quark-hadron matter EoS.
The final hybrid EoS is comprised of two separate RDF EoS for each phase of matter, combined by utilizing a Maxwell phase transition construction.

The correct thermodynamic treatment of quark matter depends on the physical constraints of the system which are to be probed.
For large volumes of quark matter that exist for long times, i.e. the thermodynamic limit, the conserved quantities are quark number (or baryon number), electron charge, color charge, and lepton number.
Each of these conserved quantities has an associated chemical potential. These large systems must remain charge- and color-neutral, which determines the values for $\mu_{Q}$ and $\mu_{\textrm{C}}$. This leaves a three-dimensional phase-space parametrized by the baryon chemical potential ($\mu_{\textrm{B}}$), the lepton chemical potential ($\mu_{\textrm{l}}$), and the temperature ($T$).

\subsection{Relativistic density-functional approach}
\label{ss:iqg}

The RDF approach to interacting quark matter can be obtained using the path integral approach to the partition function, which analogous to the treatment of the Walecka model of nuclear matter in
\cite{Kapusta:1989tk} takes the form
\begin{eqnarray}
	\label{Z}
	\mathcal{Z} = \int \mathcal{D}\bar{q}\mathcal{D}q \exp\left\{\int_0^\beta d\tau\int_Vd^3 x
	\left[\mathcal{L}_{\rm eff} + \bar{q}\gamma_0 \hat{\mu} q\right]\right\}~,
\end{eqnarray}
where in the two-flavor case
\begin{eqnarray}
	q=\left(\begin{array}{c}q_u\\q_d\end{array}\right),
\end{eqnarray}
and $\hat{\mu}={\rm diag}(\mu_u,\mu_d)$ is the diagonal matrix of the chemical potentials conjugate to the conserved numbers of up and down quarks.
The effective Lagrangian density is given by
\begin{eqnarray}
	\label{Leff}
	\mathcal{L}_{\textrm{eff}}&=& \mathcal{L}_{\textrm{free}}
	- U(\bar{q}q, \bar{q}\gamma_0q)~,\\
	\mathcal{L}_{\textrm{free}}&=&\bar{q}\left(-\gamma_0 \frac{\partial}{\partial \tau}
	+ i \vec{\gamma}\cdot \vec{\nabla} - \hat{m}\right) q~,
\end{eqnarray}
where $\hat{m}={\rm diag}(m_u,m_d)$ is the matrix of current quark masses.
The interaction is given by the potential energy density $U(\bar{q}q, \bar{q}\gamma_0q)$ which in general is a nonlinear functional  of the field representations of the scalar and vector quark densities (in the isotropic case, the vector four-current reduces to its zeroth component).
To achieve a quasiparticle representation, the potential energy density shall depend linearly on the Dirac spinor bilinears representing the relevant currents of the system.

The linearization of the interaction is facilitated by a Taylor expansion around the corresponding expectation values
\begin{eqnarray}
	\label{ns}
	{\left<\bar{q}q\right>} &=& n_{\rm s} = \sum_{f=u,d} n_{{\rm s},f}
	=-\sum_{f=u,d}\frac{T}{V}\frac{\partial }{\partial m_f} \ln \mathcal{Z}~,
	\\
	{\left<\bar{q}\gamma_0q\right>} &=& n_{\rm v} =\sum_{f=u,d} n_{{\rm v},f}
	= \sum_{f=u,d} \frac{T}{V}\frac{\partial }{\partial \mu_f} \ln \mathcal{Z}~,
	\label{nv}
\end{eqnarray}
of the scalar density $n_{\rm s}$ and the vector density $n_{\rm v}$, respectively.
This expansion results in
\begin{eqnarray}
	U(\bar{q}q,\, \bar{q}\gamma_0q) &=&
	U(n_{\rm s}, n_{\rm v})
	+ (\bar{q}q -  n_{\rm s})\Sigma_\mathrm {s}
	\nonumber\\ &&
	+ (\bar{q}\gamma_0q -  n_{\rm v}) \Sigma_\mathrm {v} +\ldots~,
\end{eqnarray}
where we have introduced the notation
\begin{eqnarray}
	\Sigma_\mathrm {s} &=& \frac{\partial U(\bar{q}q, \bar{q}\gamma_0q)}{\partial (\bar{q}q)}
	\bigg|_{\bar{q}q = n_{\rm s}} = \frac{\partial U(n_{\rm s}, n_{\rm v})}{\partial n_{\rm s}} \;,\\
	\Sigma_\mathrm {v} &=& \frac{\partial U(\bar{q}q, \bar{q}\gamma_0q)}{\partial (\bar{q}\gamma_0q)} \bigg|_{\bar{q}\gamma_0q = n_{\rm v}} = \frac{\partial U(n_{\rm s}, n_{\rm v})}{\partial n_{\rm v}}\;\textrm{,}
\end{eqnarray}
for the scalar and vector selfenergies, respectively.
The expansion is truncated at the second term, assuming the fluctuations about the expectations of the fields are small.
Employing this quasiparticle approximation for the effective Lagrangian (\ref{Leff}), the partition function (\ref{Z}) takes the form
\begin{eqnarray}
	\label{Zeff}
	\mathcal{Z} = \int \mathcal{D}\bar{q}\mathcal{D}q \exp\left\{\mathcal{S}_{\rm quasi}[\bar{q},q] -\beta V \Theta[n_{\rm s},n_{\rm v}]\right\}~,
\end{eqnarray}
where the quasiparticle action in Fourier-Matsubara representation is given by \cite{Kapusta:1989tk}
\begin{eqnarray}
	\mathcal{S}_{\rm quasi}[\bar{q},q] &=&\beta\sum_n\sum_{\vec{p}} \bar{q}\, G^{-1}(\omega_n,\vec{p})\, q~,
	\\
	G^{-1}(\omega_n,\vec{p}) &=& \gamma_0(-i\omega_n + \hat{\mu}^*) -\vec{\gamma}\cdot\vec{p} - \hat{m}^*~,
\end{eqnarray}
with the Dirac effective mass matrix $\hat{m}^*=\hat{m}+\Sigma_{\rm s}$, the renormalized chemical potential matrix $\hat{\mu}^*=\hat{\mu}-\Sigma_{\rm v}$, and the effective potential energy density
\begin{equation}
	\Theta[n_{\rm s},n_{\rm v}] = U(n_{\rm s},n_{\rm v}) - \Sigma_{\rm s} n_{\rm s} - \Sigma_{\rm v} n_{\rm v}~.
\end{equation}
The functional integral can be performed in this quasiparticle approximation with the result
\begin{eqnarray}
	\label{Zquasi}
	\mathcal{Z}_{\rm quasi} = \int \mathcal{D}\bar{q}\mathcal{D}q \exp\left\{\mathcal{S}_{\rm quasi}[\bar{q},q]\right\}=\det[\beta G^{-1}]~,
\end{eqnarray}
where the determinant operation acts in momentum-frequency space as well as on the Dirac, flavor, and color indices.
Using the identity $\ln \det A = {\rm Tr} \ln A$ and the representation of the gamma matrices, one obtains for the pressure of the homogeneous system
(for details see, e.g., Ref.~\cite{Kapusta:1989tk})
\begin{eqnarray}
	P_{\rm quasi}&=&\frac{T}{V}\ln \mathcal{Z}_{\rm quasi} = \frac{T}{V} {\rm Tr} \ln [\beta G^{-1}]\\
	&=&2N_c \sum_{f=u,d} \int \frac{d^3 p}{(2\pi)^3}
	\left\{
	 T\ln \left[1+{\rm e}^{-\beta (E^*_f-\mu^*_f)} \right]
	\right.\nonumber\\&&\left.
	 + T\ln \left[1+{\rm e}^{-\beta(E^*_f+\mu^*_f)} \right]
	\right\}~,
	\label{Pquasi}
\end{eqnarray}
where we have tacitly used the so-called ``no sea'' approximation (as is customary in the Walecka model) by removing the vacuum energy term which corresponds to the phase space integral over the one-particle energy $E^*_f=\sqrt{p^2+{m^*_f}^2}$.
After partial integration the quasiparticle contribution (\ref{Pquasi}) to the pressure takes the form ($N_c=3$)
\begin{eqnarray}
	P_{\rm quasi}= \sum_{f=u,d} \int \frac{dp}{\pi^2} \frac{p^4}{E^*_f} \left[f(E^*_f-\mu^*_f) + f(E^*_f+\mu^*_f)\right] ~,
\label{Pquasi2}
\end{eqnarray}
where $f(E)=1/[1+\exp (\beta E)] $ is the Fermi function which at $T\to 0$ degenerates to a step function.
In this $T=0$ case which is relevant for compact stars, the total pressure becomes
\begin{eqnarray}
\label{Ptot}
P&=& \sum_{f=u,d} \int_0^{p_{{\rm F},f}} \frac{dp}{\pi^2} \frac{p^4}{E^*_f} -  \Theta[n_{\rm s},n_{\rm v}] ~,
\end{eqnarray}
with the Fermi momentum of the quark flavor $f$ defined as
\begin{equation}
	p_{{\rm F},f}=\sqrt{{\mu^*_f}^2-{m^*_f}^2}\;.
\end{equation}
In order to evaluate the thermodynamics of the RDF approach, we have to solve a self-consistency problem, since the pressure is a functional of the scalar and vector densities which themselves are defined as derivatives of the pressure by Eqs.~(\ref{ns}) and (\ref{nv}), which now take the form
\begin{eqnarray}
	n_{\rm s} &=&- \sum_{f=u,d} \frac{\partial P}{\partial m_f}
	= \frac{3}{\pi^2}\sum_{f=u,d} \int_0^{p_{{\rm F},f}} dpp^2  \frac{m^*_f}{E^*_f} \;,
\\
n_{\rm v} &=& \sum_{f=u,d} \frac{\partial P}{\partial \mu_f}
	= \frac{3}{\pi^2}\sum_{f=u,d} \int_0^{p_{{\rm F},f}} dpp^2 = \frac{p_{{\rm F},u}^3+p_{{\rm F},d}^3}{\pi^2}\;.
	\nonumber\\
\end{eqnarray}

The general RDF approach to quark matter as outlined here is open to applications for a variety of interaction potentials, in both the scalar and vector channels.
The general form of the function $U(n_{\rm s},n_{\rm v})$ can be modeled after specific interactions.
In the next section, the specific interaction potential for the present RDF quark-matter model is introduced and the resulting selfenergies $\Sigma_{\rm s}$ and $\Sigma_{\rm v}$ are discussed.

\subsection{Density functional for quark matter}

\begin{figure}[!h]
	\vspace{-5mm}
	\includegraphics[scale=0.45]{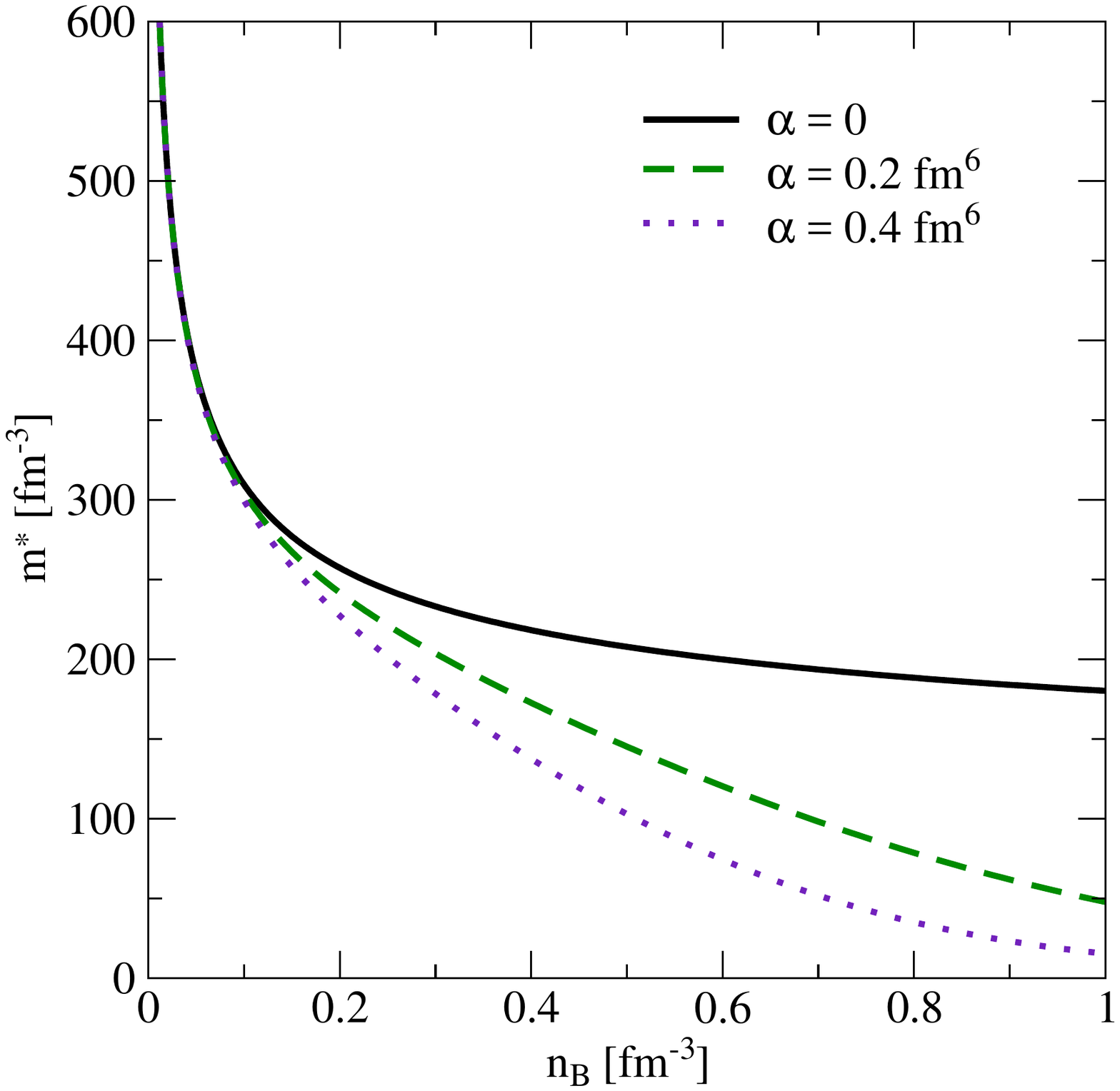}
	\vspace{-8mm}
	\caption{Effective quark mass as a function of the baryon density.
	The effective confinement manifests itself by a divergence of the quasiparticle mass at low densities.}
	\label{fig:mass_vs_nb}
\end{figure}

When the RDF quasiparticle Fermi gas was derived, the interaction potential in the Lagrangian was introduced as a functional of the quark currents in the scalar and vector channels,
${\bar{q}q}$ and ${\bar{q}\gamma_0 q}$, respectively.
This potential was expanded about the mean-field values, $\left<\bar{q}q\right> = n_{\rm s}$ and
$\left<\bar{q}\gamma_0 q\right> = n_{\rm v}$, utilizing the Taylor expansion to the first-order.
The Taylor expansion coefficients could then be absorbed in the definition of the fermionic quasiparticle properties, i.e., the effective masses $m^*_f$ and the effective chemical potentials $\mu^*_f$.
As a result, the mean-field thermodynamic potential potential could be given in a thermodynamically consistent form, just based on the knowledge of the interaction potential, now given in terms of the scalar and vector densities.
In the following, we use flavor independent quark masses $m=m_u=m_d$ entailing
$\hat{m}^*={\rm diag}(m^*,m^*)$ with $m^*=m+\Sigma_{\rm s}$.
For the density functional of the interaction energy, we adopt in the present work the form
\begin{equation}
\label{potential}
	U({n}_\mathrm s,{n}_\mathrm v) = D({n}_\mathrm v){n}_\mathrm s^{2/3} +a {n}_\mathrm v^2 + \frac{b {n}_\mathrm v^4}{1+c {n}_\mathrm v^2}\;\textrm{,}
\end{equation}
Let us explain the terms occurring in (\ref{potential}) more in detail.
The first term captures aspects of (quark) confinement through the density dependent
scalar selfenergy
\begin{eqnarray}
\label{eqnshifts2}
	\Sigma_\mathrm s &=& \frac{2}{3}D({n}_\mathrm v){n}_\mathrm s^{-1/3}~,
\end{eqnarray}
defining the effective quark mass $m^*$.
The effective mass diverges\footnote{For color neutral hadrons this divergence of the selfenergy is entirely compensated by that of the confining interaction in the equation of motion \cite{Glozman2009}.}
for densities approaching zero, see Fig.~\ref{fig:mass_vs_nb}, and thus suppresses the occurrence of the quasiparticle excitations corresponding to these degrees of freedom.
For quark matter in compact stars, such a mechanism has recently been used in Ref.~\cite{Li:2015ida}.
Note that in its nonrelativistic formulation with energy shifts \cite{Ropke:1986qs}, the SFM has already been applied successfully to describe massive hybrid stars with quark-matter cores \cite{Blaschke:1989nn}.

The second term in (\ref{potential}) stands for the repulsion stemming from a four-fermion interaction in the Dirac vector channel while the last term is a higher order (8-quark) repulsive interaction in the vector channel.
Such higher order vector meanfields have been considered already in the description of nuclear matter (see, e.g., Ref.~\cite{Serot:1997xg}), and it is therefore natural to invoke them also in the description at the quark level.
The higher order quark interactions have been introduced in \cite{Benic:2014iaa} for the description of hybrid stars in order to provide a sufficient stiffening at high densities required to fulfill the $2M_\odot$ mass constraint from the precise mass measurement of  \cite{Demorest:2010bx} and \cite{Antoniadis:2013pzd}.
This allows one to obtain a separate third family of high-mass hybrid stars \cite{Benic:2014jia}.
The denominator in the last term of (\ref{potential}) makes sure that for appropriate choice of the parameters $b$ and $c$ the speed of sound $c_\mathrm s=\sqrt{\partial P/\partial \varepsilon}$ does not exceed the speed of light.
All terms in the density functional (\ref{potential}) that contain the vector density contribute to the shift defining the effective chemical potentials $\mu^*_f$,
\begin{eqnarray}
	\Sigma_\mathrm v = 2 a n_\mathrm v+\frac{4b n_\mathrm v^3}{1+c n_\mathrm v^2}
	- \frac{2 b c n_\mathrm v^5}{(1+c n_\mathrm v^2)^2} + \frac{\partial D(n_\mathrm v)}{\partial n_\mathrm v}{n}_s^{2/3}\,\textrm{.}
	\label{mu-shift}
	\end{eqnarray}
Note that the last term in (\ref{mu-shift}) comes from a vector density dependence in the effective string tension which is a new aspect of the present approach and a generalization of the SFM.
%
%

\begin{figure}[!h]
	\includegraphics[scale=0.260]{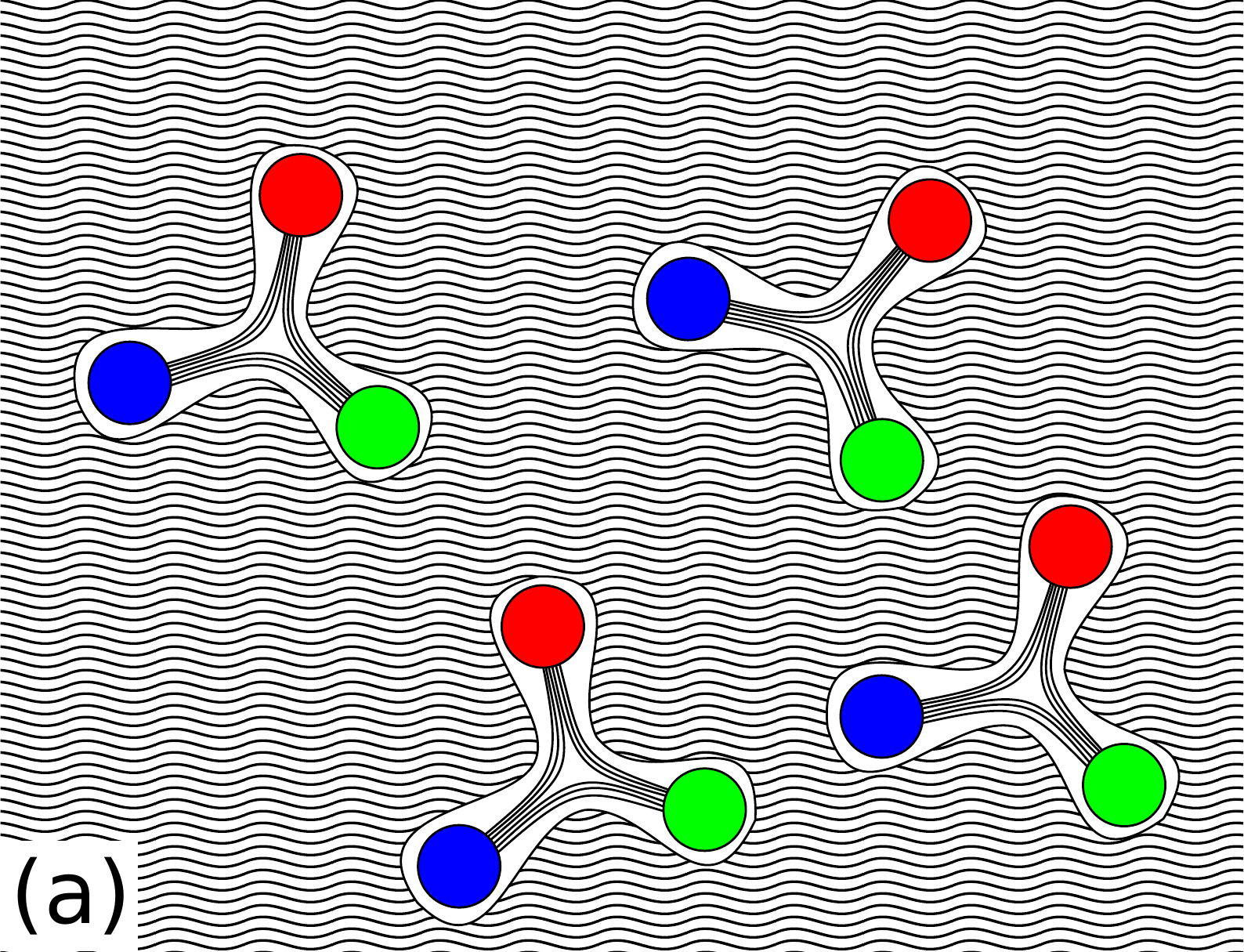} 
	\includegraphics[scale=0.260]{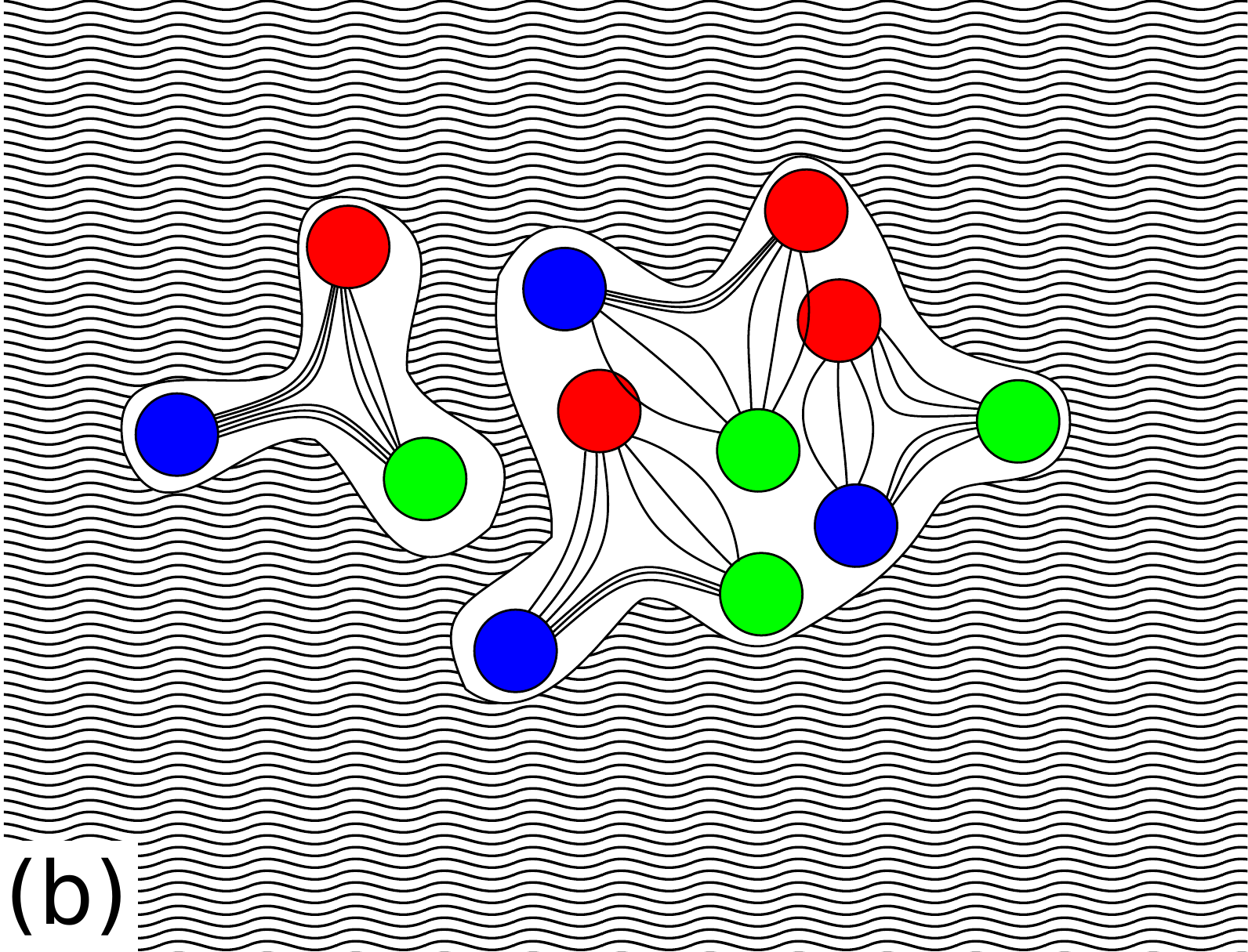}
	\caption{Illustration of the effective reduction of the string tension (density of color field lines) at high densities. At low densities (a) the field lines are compressed to thin flux tubes by the dual Meissner effect while at high densities (b) this pressure is reduced and consequently the effective string tension is lowered.}
	\label{fig:meissner}
\end{figure}

The SFM modification takes into account the occupation of the surrounding medium by color fields, which lead to  an effective reduction of the in-medium string tension.
It is accounted for by multiplying the vacuum string tension $D_0$ between color charges with the available volume fraction (\ref{eq:hadron_avol})
\begin{eqnarray}
\label{Dv}
	D(n_\mathrm v) &=& D_0 \Phi(n_\mathrm v)~.
\end{eqnarray}
This reduction of the string tension is understood as a consequence of a modification of the pressure on the color field lines by the dual Meissner effect since the reduction of the available volume corresponds to a reduction of the nonperturbative dual superconductor QCD vacuum that determines the strength of the confining potential between the quarks.
We illustrate this situation in Fig.~\ref{fig:meissner}.

Taking into account all of the mean-field contributions (\ref{potential})-(\ref{Dv}) discussed above,
the quark-matter EoS of the RDF approach
\begin{eqnarray}
P=P_{\rm quasi} - \Theta[n_{\rm s},n_{\rm v}]
\end{eqnarray}
is defined for the SFM model and will be employed in the following for the $T=0$ case (\ref{Ptot}) that is relevant for applications to compact stars.
\subsection{Hadronic phase}

For the hadronic phase we use the well known DD2 model by Typel et al. \cite{Typel:2005ba}
with the parameterization given in \cite{Typel:2009sy}.
It is a relativistic density functional model, based on meson exchange interactions.
It describes the properties of nuclear matter at saturation density and below very well, also in accordance with the chiral EFT approach \cite{Kruger:2013kua}, see also \cite{Fischer:2013eka}.
To improve the higher density behavior, a generalized excluded volume effect is included
according to Ref.~\cite{Typel:2016abc} using the Gaussian form of the (isospin independent) available volume fraction
\begin{equation}
\label{eq:hadron_avol}
	\Phi(n_\mathrm B) = \begin{cases}
					1, & \mathrm{if ~}n_\mathrm{B}<n_0 \\
					e^{- \alpha (n_\mathrm{B} - n_0)^2}, &  \mathrm{if ~}n_\mathrm{B}>n_0
				\end{cases}\;\mathrm{,}
\end{equation}
where $\alpha=v |v| / 2$ and $v$ is the excluded volume parameter  that will be varied in our current work
to investigate its effect on the neutron star mass-radius curves.
The available volume fraction is coupled to the degeneracy factor creating $g^{(\mathrm{eff})}_i= g_i \Phi_i$. This effectively reduces the degrees of freedom as the system becomes more dense.

\subsection{Compact stars}

Compact stars are large systems of degenerate matter, which tend to exist for quite long times, so the thermodynamic limit constraints are valid for describing these systems.
Neutrinos are considered capable of escaping from cold compact stars due to their negligible interaction which entails a sufficiently large mean free path, so lepton number is not conserved.
This leaves compact star matter to be constrained by baryon number and $T$. For our further discussion, we shall be operating in the zero-temperature regime, further reducing the parameters to just the baryon number.
For quark matter in compact star systems, due to the large threshold chemical potential for strange quarks $\sim\!1500$ MeV (and larger for the heavier flavors of quarks), contributions from heavy quarks are negligible  \cite{Kettner1995} and therefore not considered in this approach.
Utilizing the above mentioned assumptions, both the hadronic and quark-matter phase in compact stars obey the following conditions. The matter is in $\beta$-equilibrium with electrons,
\begin{eqnarray}
&&	d \to u + e^{-} + \bar{\nu}_{e} ~,~~ u + e^{-} \to d + {\nu}_{e}\\
&&	n \to p + e^{-} + \bar{\nu}_{e} ~,~~ p + e^{-} \to n + {\nu}_{e} \;\textrm{.}
\end{eqnarray}

The relevant chemical potentials $\mu_u$, $\mu_d$, and $\mu_e$  ($\mu_p$, $\mu_n$, and $\mu_e$ for the hadronic phase) satisfy the weak equilibrium ($\beta$-equilibrium) condition
condition
\begin{eqnarray}
&&	\mu_d - \mu_u = \mu_e\;\textrm{,}\\
&&	\mu_n - \mu_p = \mu_e\;\textrm{,}
\end{eqnarray}
where we have assumed that the neutrinos escape without interacting\footnote{This case applies when the compact star is sufficiently cooled down below the neutrino opacity temperature $\sim 1$ MeV, so that the radius of the neutrino sphere exceeds that of the star and neutrinos leave the star, not participating in the
$\beta-$equilibrium processes.}. The quark and baryon densities are related by
\begin{equation}
	n_\mathrm{B} = \frac{1}{3} n_\mathrm{v}=\frac{1}{3} (n_{{\rm v},u} + n_{{\rm v},d})\;\textrm{.}
\end{equation}
The charge density (without muons)  of the quark-matter phase is
\begin{equation}
	n_{\mathrm{Q},q} = \frac{2}{3} n_{{\rm v},u} - \frac{1}{3} n_{{\rm v},d} - n_{e}~,
\end{equation}
while for the hadronic phase it is
\begin{equation}
	n_{\mathrm{Q},h} = n_{p} - n_{e}~.
\end{equation}
Electric charge neutrality of the homogeneous phases requires $n_{\mathrm{Q},q}=n_{\mathrm{Q},h}=0$.
In the case of a phase transition, different possibilities to realize global charge neutrality occur.

\subsection{Phase transition}

Generally, a mixed phase can occur when constructing the quark-hadron phase transition, e.g., by adjusting a volume fraction $\chi$ of the quark phase so that charge neutrality in that phase is
fulfilled globally \cite{Glendenning:1992vb}
\begin{equation}
0=n_\mathrm{Q}=\chi n_{\mathrm{Q},q} + (1-\chi) n_{\mathrm{Q},h}~.
\end{equation}
In that situation the quark and hadron phases may separately be charged.
A more sophisticated construction of the phase transition considers the occurrence of structures (so-called ``pasta phases'') with an interplay of surface tension, Coulomb energy and charge screening effects.
For a recent work see, e.g., Ref.~\cite{Yasutake:2014oxa} and references therein.

In the present work, we shall use a simple Maxwell construction for the phase transition between two separately charge neutral phases, requiring $n_{\mathrm{Q},q}=n_{\mathrm{Q},h}=0$.

The two distinct phases are then matched by assuming chemical ($\mu$), thermal ($T$), and mechanical ($P$) equilibrium at the phase-transition, outside the phase-transition the phase with higher pressure (lower grand canonical potential), is the one chosen.
Technically, we plot the isotherms (${T^\mathrm H = T^\mathrm Q = 0}$) of both phases in pressure over baryon chemical potential and merge them at the crossing point (where ${\mu_B^\mathrm H = \mu_B^\mathrm Q}$ and ${P^ \mathrm H = P^\mathrm Q}$ is fullfilled).
In this way thermodynamical consistency is fulfilled, because temperature and chemical potentials are the natural variables of the grand potential or the pressure (${\Omega(T,\{\mu_i\}) = -P(T,\{\mu_i\})}$).

\section{Results}
\label{Results}

Theoretical studies of the properties of compact stars have increased in prevalence in recent years because of the growing amount of observational data that has been collected for the masses and radii of pulsars. The measurements are used to constrain and elevate the most promising EoSs, while eliminating the others. The constraints imposed by the observational data provide a metaphorical `playground' for probing properties of matter unattainable in terrestrial experiments. This `playground' allows theorists to investigate different phenomenology which can explain the data, such as the investigation into a hadron to quark-matter phase transition.

By solving the Tolman-Oppenheimer-Volkoff equations using the EoS designed from the RMF quasiparticle Fermi gas, a unique type of mass-radius (M-R) relation is obtained. This relation is altered in distinct ways by varying each individual parameter. The four parameters to be systematically varied, while observing their effects on the observable M-R relation, are the hadronic excluded volume parameter
$\alpha=v|v|$ in Eq.~(\ref{eq:hadron_avol}); the high-density, 8-quark collision coupling parameter
$b$; the transition-density, 4-quark collision coupling parameter $a$; and the available volume fraction $\Phi$, which is dependent on the parameter $\alpha$.

\begin{figure}[!thb]
\vspace{-10mm}
	\includegraphics[scale=0.45]{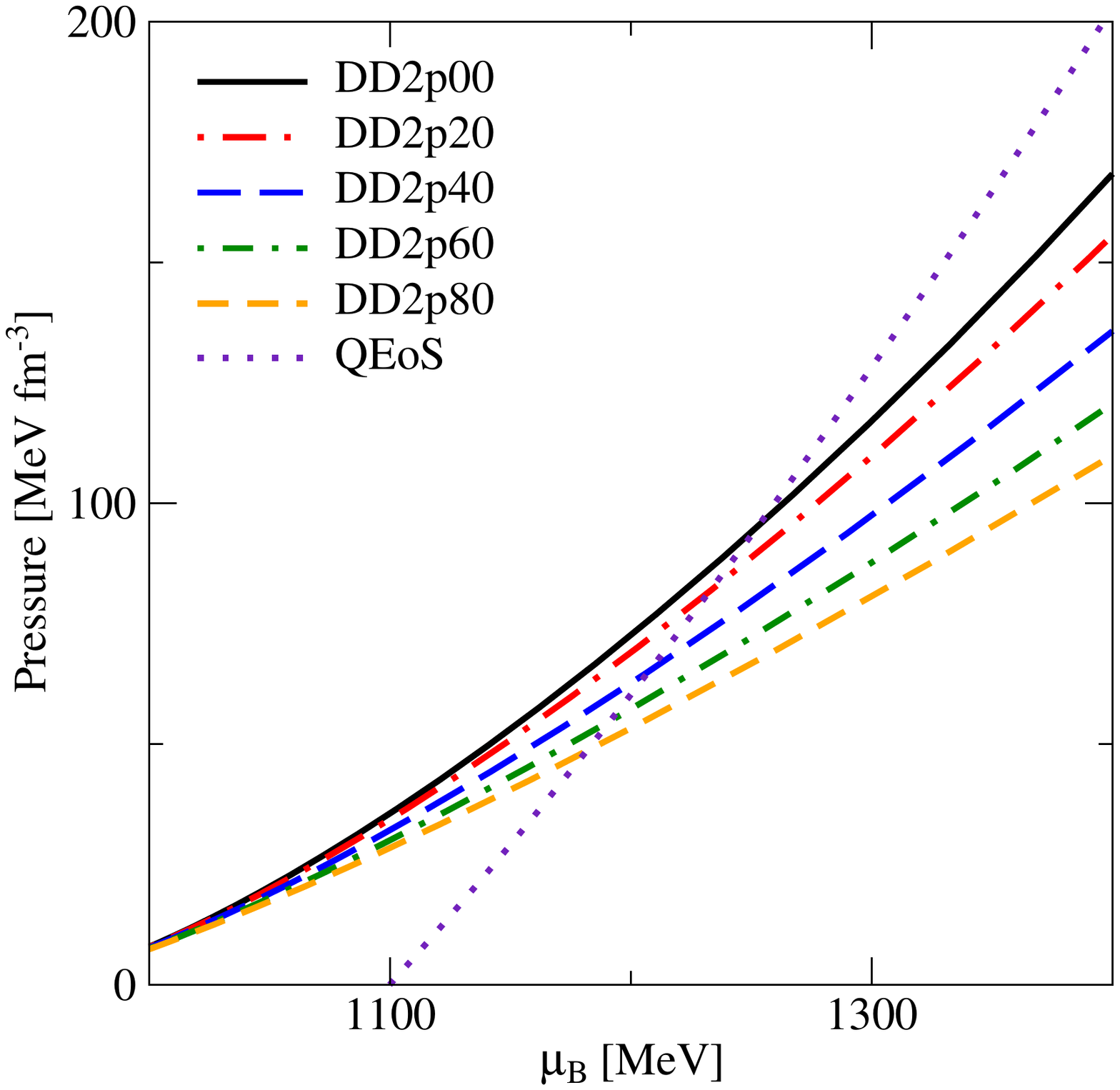}
\vspace{-10mm}
	\caption{(Color online) Pressure versus baryon chemical potential for the hadronic DD2 EoS with different excluded volume parameters 
	and for a quark-matter EoS (QEoS) with parameters $\alpha = 0.2$ fm$^6$, $a = -2.0 \mathrm{\;MeV\;fm^3}$, $b=2.0 \mathrm{\;MeV\;fm^9}$, and $c=0.036 \mathrm{\;fm^6}$.
	As the excluded volume is increased, the slope of the hadronic DD2 is lowered and the onset of the hadron-to-quark-matter transition (the crossing of quark and hadron EoS) is lowered.}
	\label{fig:eos-mu_crossing}
\end{figure}

\begin{figure}[!th]
\vspace{-40mm}
	\includegraphics[scale=0.45]{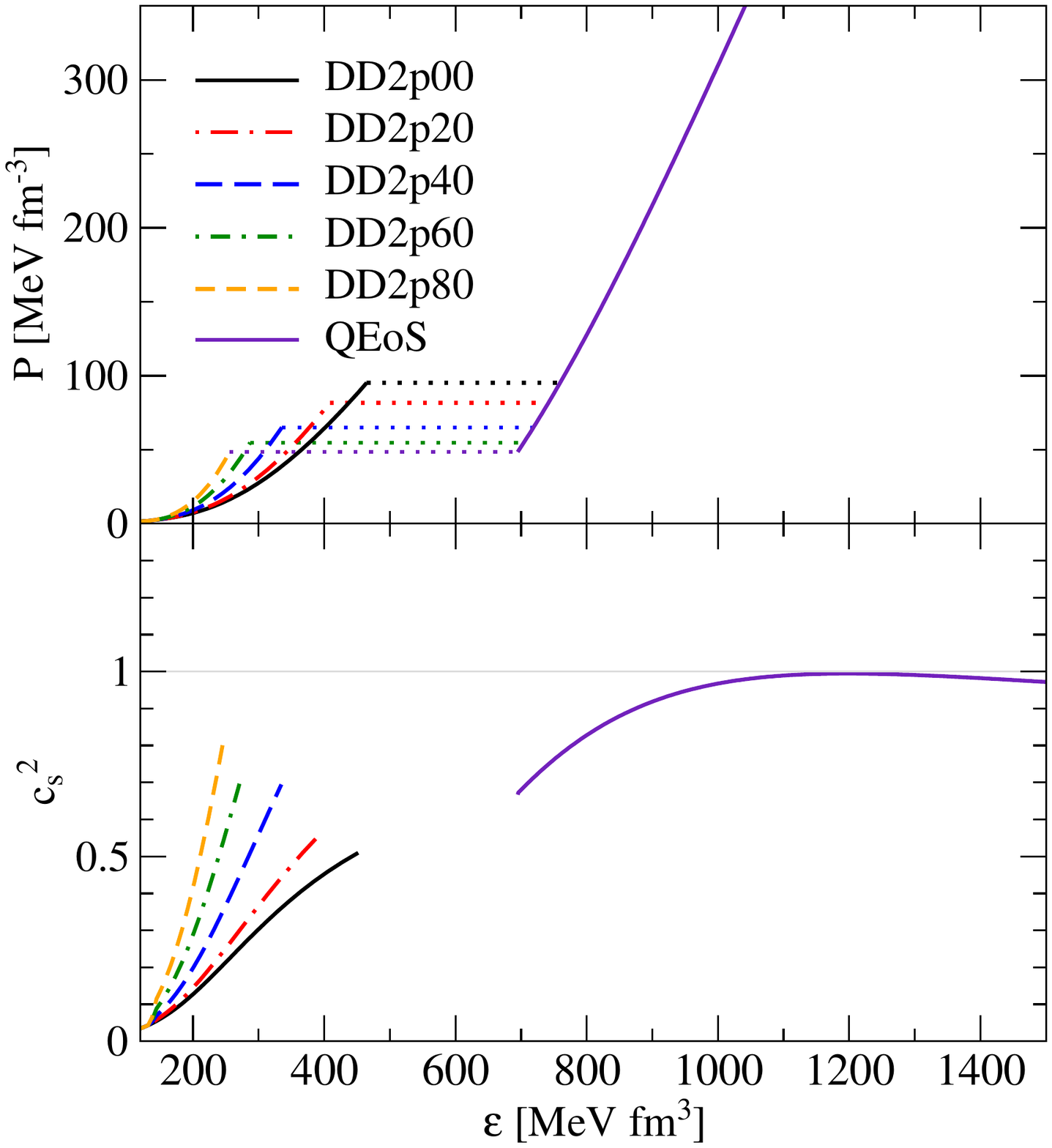}
\vspace{-10mm}
	\caption{(Color online) Upper panel: Pressure versus energy density for the hybrid EoS that emerge from the Maxwell constructions corresponding to Fg.~\ref{fig:eos-mu_crossing}. As the excluded volume parameter is increased, the slope in the hadronic phase is raised exhibiting the stiffening of the hadronic matter. Lower panel: Squared speed of sound as a function of energy density, verifying the causality condition ($c_s^2<1$).}
	\label{fig:eos-edens_crossing}
\end{figure}

\begin{figure}[!htb]
\vspace{-10mm}
	\includegraphics[scale=0.45]{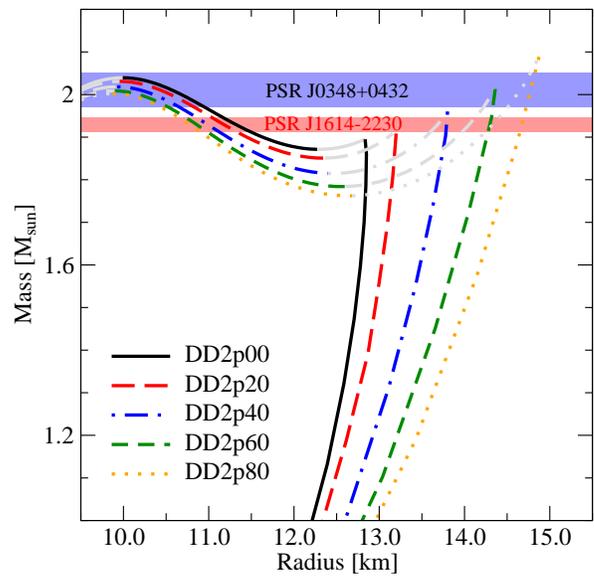}
\vspace{-10mm}
	\caption{(Color online) M-R relations for the hybrid EoS shown in
	Fig.~\ref{fig:eos-edens_crossing} with varying hadronic excluded volume parameter.
	The grey lines represent the unstable configurations of hybrid stars.}
	\label{fig:vary_exvol_TOV}
\end{figure}

It is useful to define the descriptive labels for the EoS used below. The hadronic EoS is labeled as ``DD2p\#\#", where ``p\#\#"  describes the excluded volume value used in the EoS (e.g., DD2p80 indicates the hadronic EoS has an excluded volume value of $v=8.0$ fm$^3$). For the following results, the string tension potential is determined to be $D_0^{1/2} = 240$ MeV, and the value of the parameter $c$ was set to $0.036 \mathrm{\;fm^6}$ to maintain causality for all the following parameterizations.

\subsection{Variation of the Hadronic Excluded Volume}
The first parameter to be addressed is the DD2 excluded volume. By increasing it the hadronic matter
is stiffened, reducing the compressibility of the star and leading to an increase in the radii at a given mass
on the hadronic branch of the sequences in the M-R diagram shown in Fig.~\ref{fig:vary_exvol_TOV}, before the phase transition which manifests itself with a branch of unstable configurations (grey lines).
It is a feature of the quark-matter EoS used here that it is sufficiently stiff (close to the causality limit, see
Fig.~\ref{fig:eos-edens_crossing}) to allow for stable branches of very compact hybrid stars,
forming a so-called ``third family'' of compact stars besides the first (white dwarfs) and the second one (purely hadronic neutron stars).

Increasing the excluded volume parameter in the hadronic phase stiffens the hadronic matter. This stiffening shifts the chemical potential of the phase-transition to smaller values, resulting in an earlier transition. It is interesting to note that although the mass of the star at phase transition onset increases, the pressure at these points gets lower.

\begin{figure}[!ht]
\vspace{-10mm}
	\includegraphics[scale=0.45]{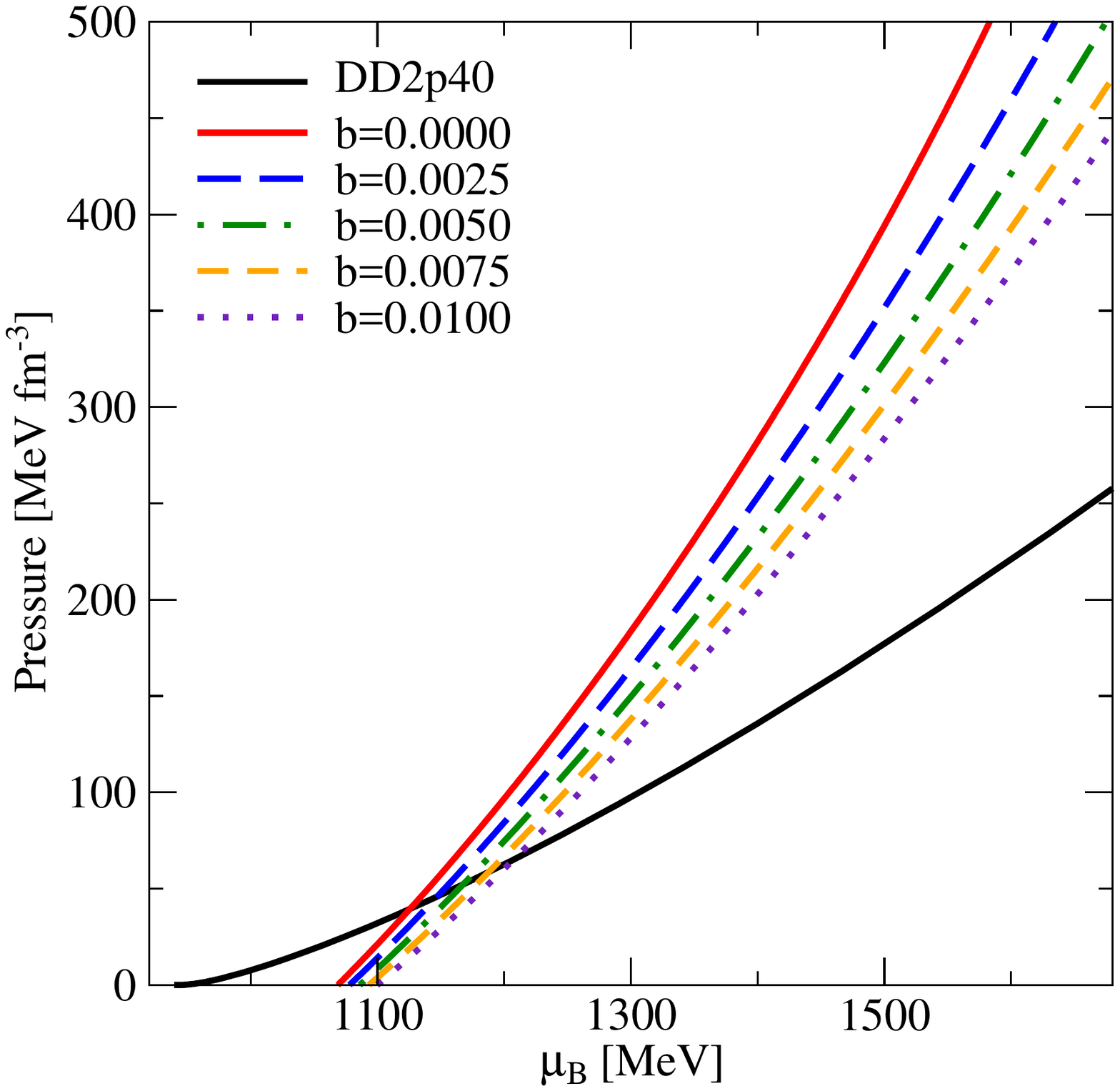}
\vspace{-10mm}
	\caption{(Color online)
	Same as Fig.~\ref{fig:eos-mu_crossing}
	but for varying values of the 8-quark coupling parameter $b$ while the other parameters are kept fixed at $\alpha = 0.2$ fm$^6$, $a = -2.0 \mathrm{\;MeV\;fm^3}$, and $c=0.036 \mathrm{\;fm^6}$, and the hadronic phase is described by the DD2p40 EoS. }
	\label{fig:eos-mu_vary_b}
\end{figure}

\begin{figure}[!htb]
\vspace{-40mm}
	\includegraphics[scale=0.45]{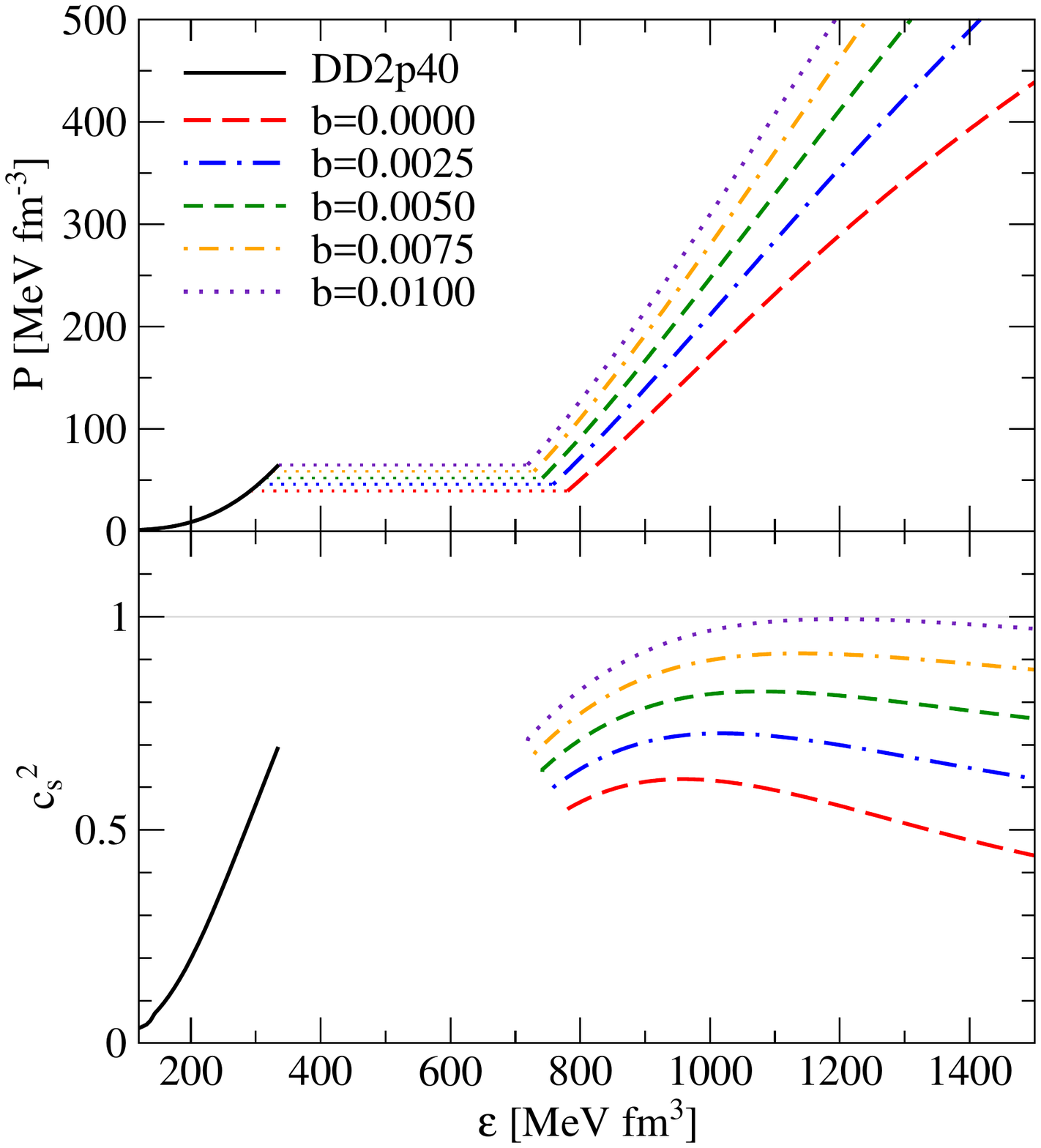}
\vspace{-10mm}
	\caption{(Color online) The same as Fig.~\ref{fig:eos-edens_crossing} for the hybrid EoS cases following from the EoS shown in Fig.~\ref{fig:eos-mu_vary_b}. }
	\label{fig:eos-edens_vary_b}
\end{figure}

\subsection{Variation of the Stiffness at High Densities}

The most influential parameter for highest densities of the quark-matter phase is the 8-quark coupling parameter $b$. From the vector shift equation (\ref{mu-shift}) which defines the modification of the quasiparticle chemical potential, the dominance of parameter $b$ at the highest densities is evident due to the coupling to the vector density to the third power. This term gives a large, repulsive contribution to the energy, i.e., an effective strong stiffening effect on the EoS, see Figs.~\ref{fig:eos-mu_vary_b} and \ref{fig:eos-edens_vary_b}.

\begin{figure}[!htb]
\vspace{-15mm}
	\includegraphics[scale=0.45]{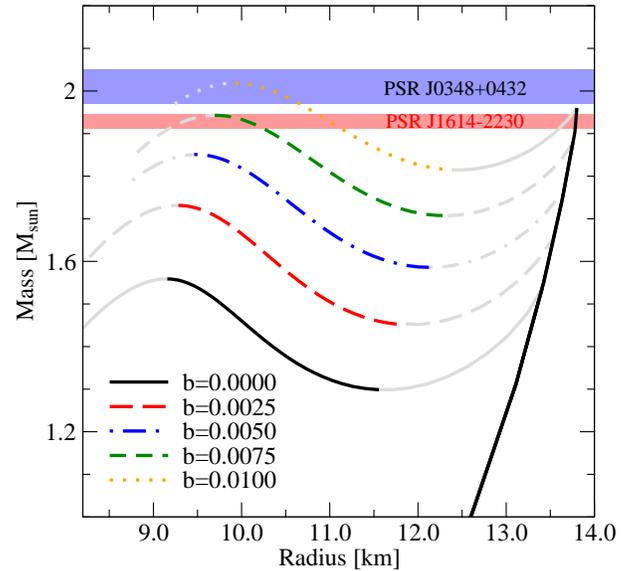}
\vspace{-10mm}
	\caption{(Color online) M-R relations for the hybrid EoS shown in Fig.~\ref{fig:eos-edens_vary_b}, differing in the values of the 8-quark coupling parameter $b$.
	}
	\label{fig:vary_b_TOV}
\end{figure}

The strongly stiffened quark-matter EoS has a substantial stiffening effect at phase-transition densities. Increasing $b$ delays the onset of quark matter to compact stars with higher mass. With increasing values of $b$, the transition from hadron to quark degrees of freedom is shifted to higher values of chemical potentials. The strong effect of the 8-quark interactions is more evident in the M-R relations, see Fig.~\ref{fig:vary_b_TOV}.

\begin{figure}[!thb]
\vspace{-13mm}
	\includegraphics[scale=0.45]{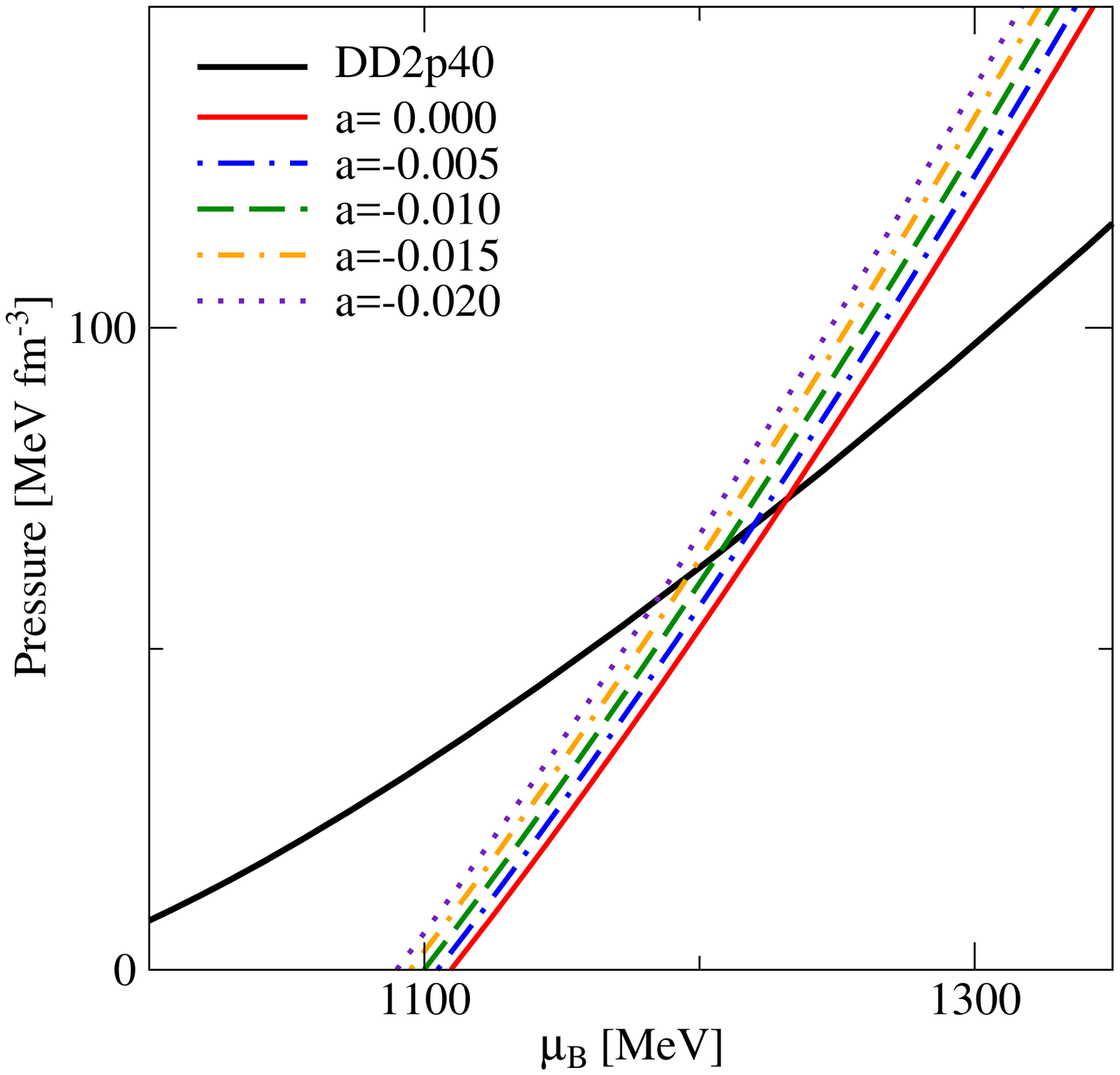}
\vspace{-10mm}
	\caption{(Color online)
	Same as Fig.~\ref{fig:eos-mu_crossing}
	but for varying values of the 4-quark coupling $a$, while remaining parameters are fixed at  $\alpha = 0.2$ fm$^6$, $b=2.0 \mathrm{\;MeV\;fm^9}$, and $c=0.036 \mathrm{\;fm^6}$.}
	\label{fig:eos-mu_vary_a}
\end{figure}

\begin{figure}[!htb]
\vspace{-40mm}
	\includegraphics[scale=0.45]{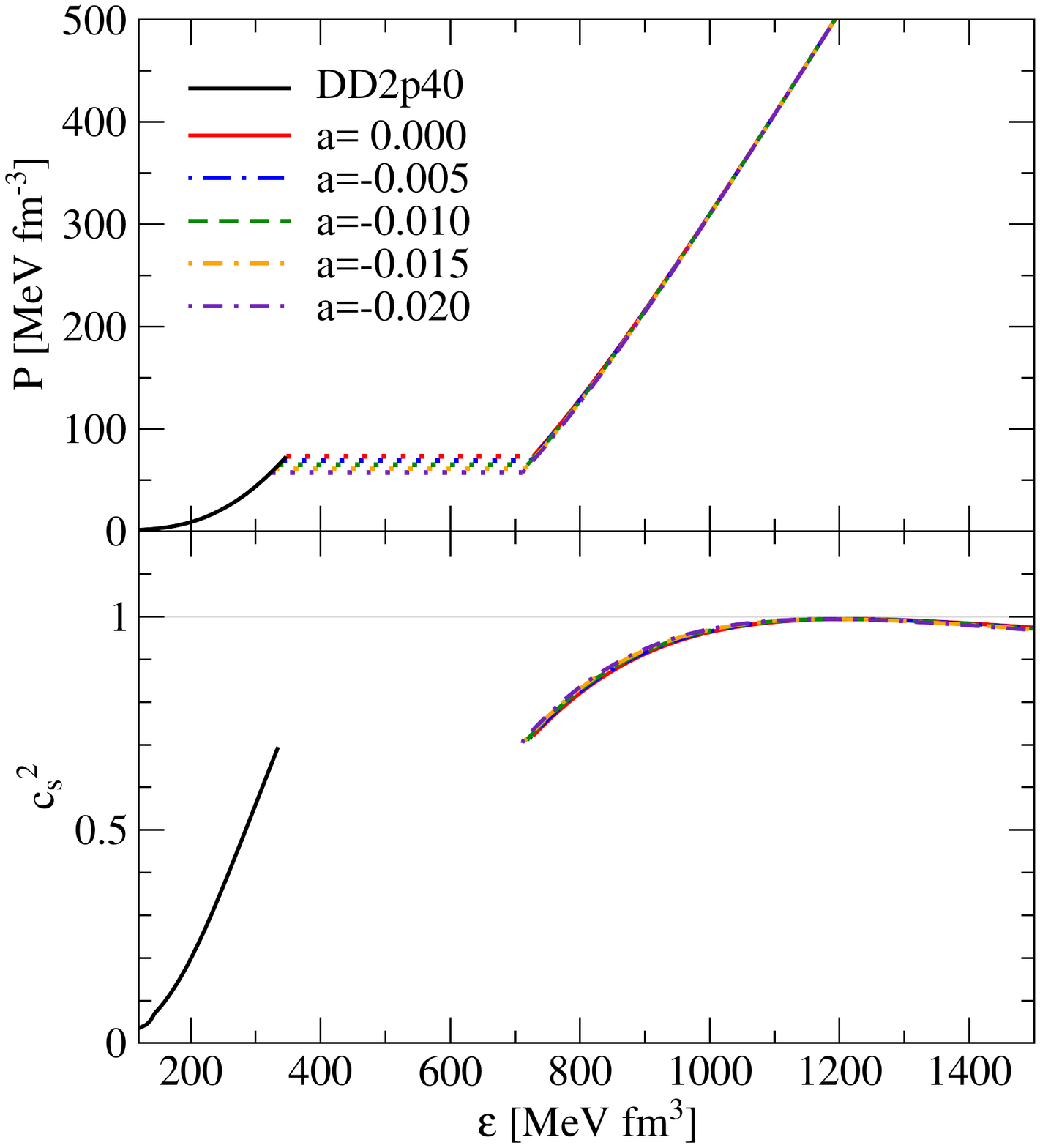}
\vspace{-10mm}
	\caption{(Color online) The same as Fig.~\ref{fig:eos-edens_vary_b} for the hybrid EoS cases following from the EoS shown in Fig.~\ref{fig:eos-mu_vary_a}. }
	\label{fig:eos-edens_vary_a}
\end{figure}

\subsection{Variation of the Stiffness at Transition Densities}

\begin{figure}[!htb]
\vspace{-10mm}
	\includegraphics[scale=0.45]{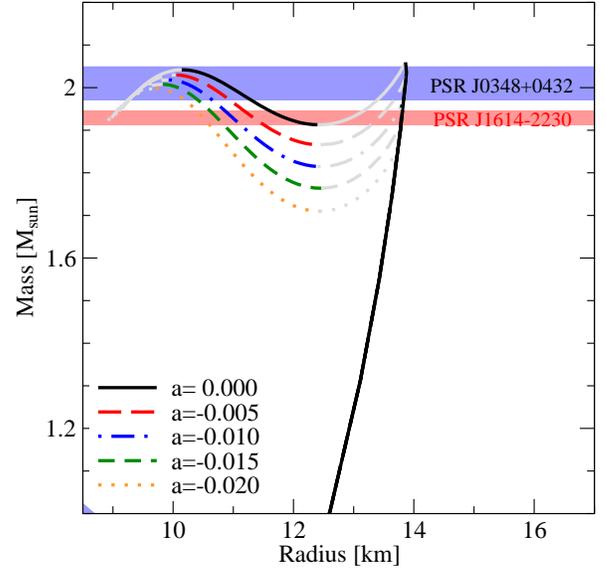}
\vspace{-10mm}
	\caption{(Color online) M-R relations for the hybrid EoS shown in Fig.~\ref{fig:eos-edens_vary_a}, differing in the values of the 4-quark coupling parameter $a$.
	}
	\label{fig:vary_a_TOV}
\end{figure}

The parameter $a$ represents a 4-quark interaction term resulting in a vector selfenergy depending linearly on the vector density.  This parameter allows for a manipulation of the properties of the phase transition point without strongly influencing on the high-density region governed by the parameter $b$, see Figs.~\ref{fig:eos-mu_vary_a}, \ref{fig:eos-edens_vary_a} and \ref{fig:vary_a_TOV}.

\begin{figure}[!htb]
\vspace{-10mm}
	\includegraphics[scale=0.45]{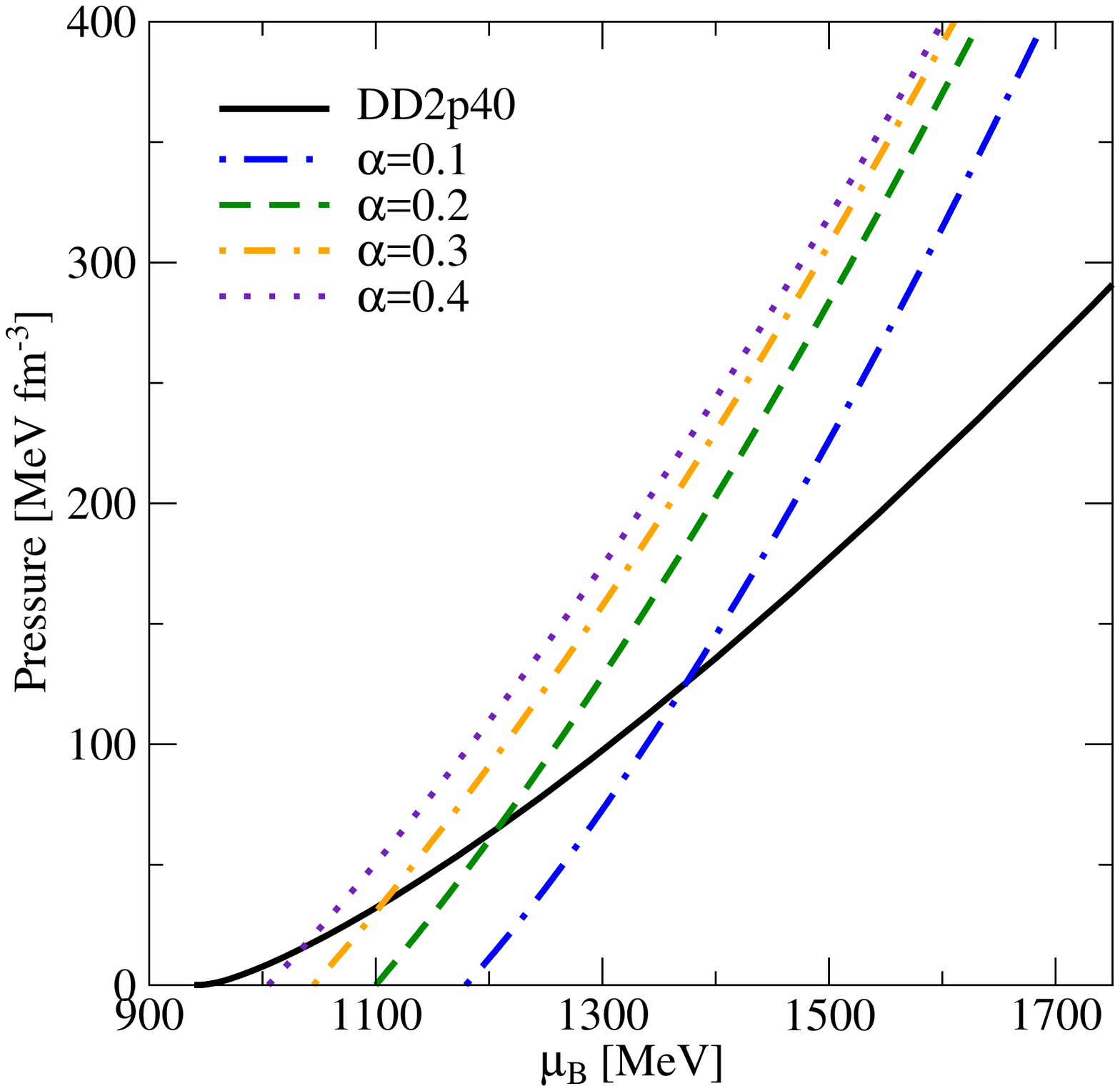}
\vspace{-10mm}
	\caption{(Color online)
	Same as Fig.~\ref{fig:eos-mu_crossing}
	for varying values of the available volume parameter $\alpha$, while the remaining parameters are fixed at $a = -2.0 \mathrm{\;MeV\;fm^3}$, $b=2.0 \mathrm{\;MeV\;fm^9}$ and $c=0.036 \mathrm{\;fm^6}$.}
	\label{fig:eos-mu_vary_avol}
\end{figure}

\subsection{Variation of the Available Volume for Quark Matter}

\begin{figure}[!htb]
	\includegraphics[scale=0.45]{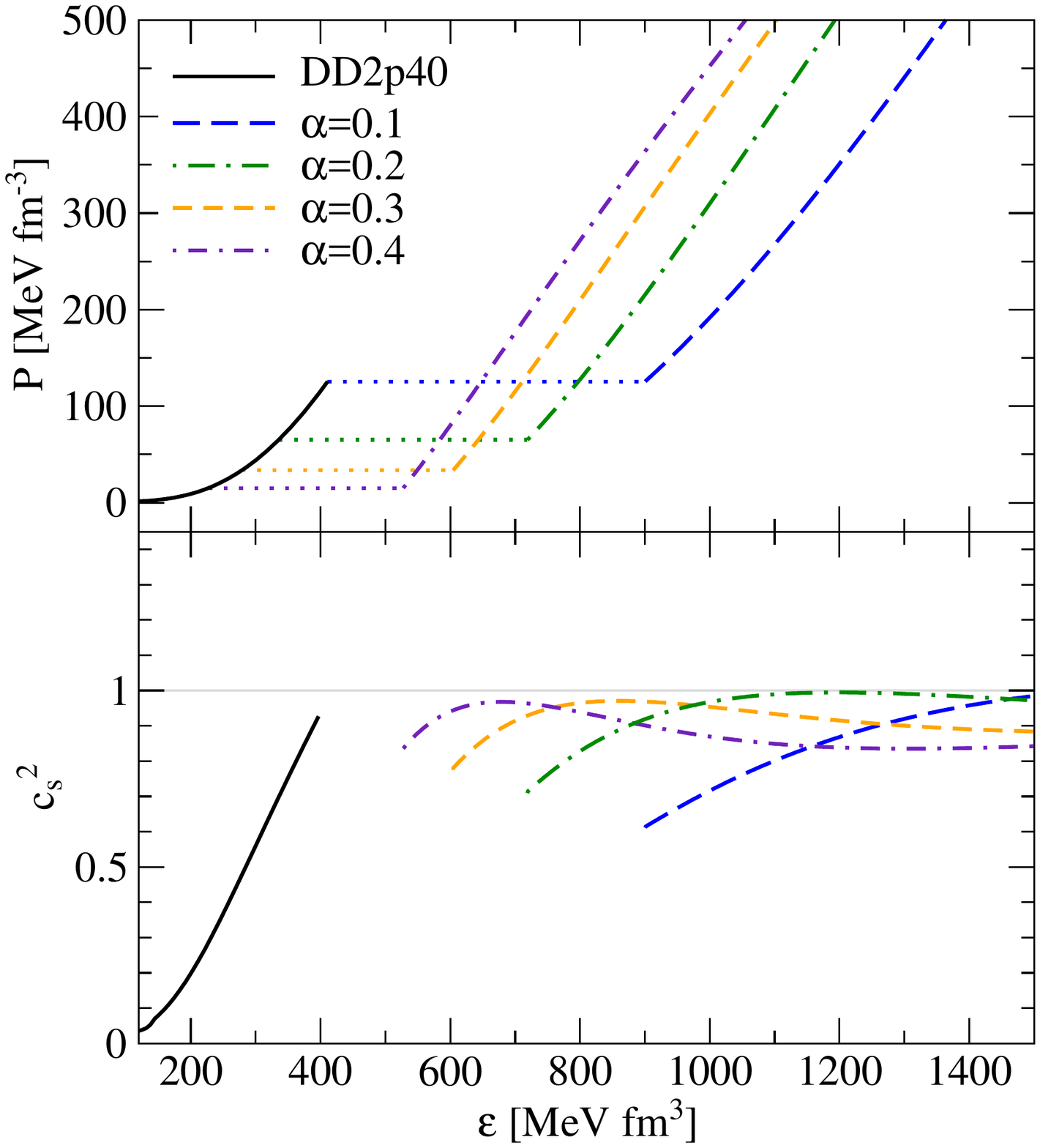}
\vspace{-10mm}
	\caption{(Color online) The same as Fig.~\ref{fig:eos-edens_vary_a} for the hybrid EoS cases following from the EoS shown in Fig.~\ref{fig:eos-mu_vary_avol}. }
	\label{fig:eos-edens_vary_avol}
\end{figure}

\begin{figure}[!th]
\vspace{-10mm}
	\includegraphics[scale=0.45]{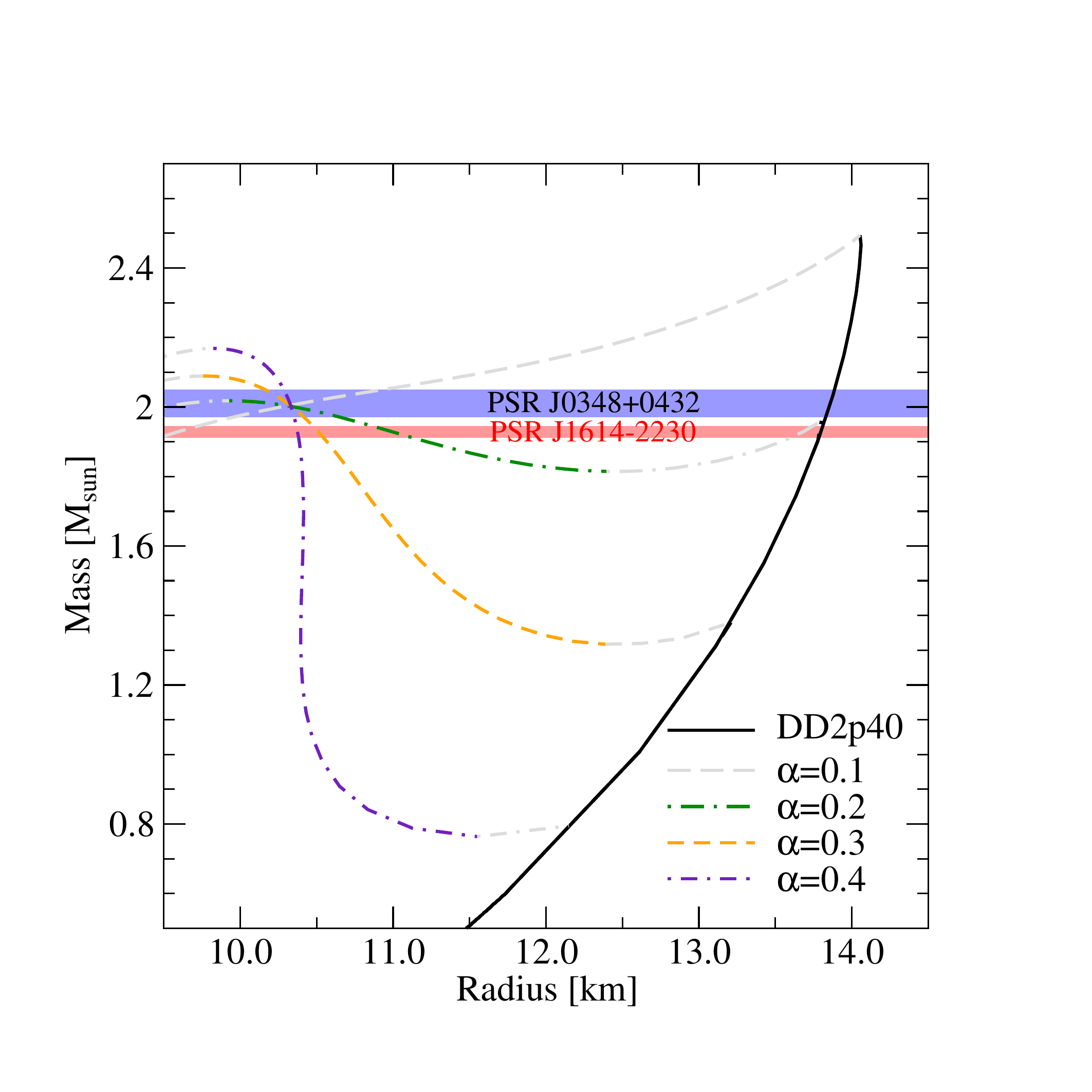}
\vspace{-10mm}
	\caption{(Color online) M-R relations for the hybrid EoS shown in
	Fig.~\ref{fig:eos-edens_vary_avol}, differing in the values of the available volume parameter
	$\alpha$.
	}
	\label{fig:vary_avol_TOV}
\end{figure}

The available volume fraction controls the scaling of the effective mass of the quasiparticle with increasing density. This fraction controls the rate at which the effective mass converges to the bare quark mass. Reducing the mass of the quasiparticle more quickly has the effect of increasing the pressure of the quark gas at lower densities. As such, the phase-transition point for this EoS is shifted down to a much lower chemical potential, see Figs.~\ref{fig:eos-mu_vary_avol} and \ref{fig:eos-edens_vary_avol}.
In Fig.~\ref{fig:vary_avol_TOV}, the strong effect of the available volume color screening can be seen.

The value of the available volume parameter $\alpha$, representing the strength of the color-screening mechanism, has a strong effect on the onset of the quark phase transition. The strength of the screening lowers the onset mass for the creation of a quark core considerably.

\section{Discussion}
\label{Discussion}

As seen in the figures presented above, the stiffness of the hadronic matter has a significant effect on the radius difference, $\Delta R_{\textrm{star}}$, of the twin configurations. As the hadronic phase is stiffened, the radii of a purely hadronic star increases. The most stiff hadronic EoS produces a neutron star of $\sim2$ M$_\odot$ and radius $\sim15.0$ km, and the resulting $\Delta R_{\textrm{star}}$ for the hybrid star twin is $\sim4.0$ km, see Fig.~\ref{fig:vary_exvol_TOV}.
At the other end of the spectrum, the hadronic EoS without any excluded volume effects has a $\Delta R_{\textrm{star}}\simeq 0.25$ km. Upcoming astronomical endeavors, such as the NICER \cite{NICER1} and SKA \cite{SKA1} observational experiments, have the possibility to resolve these differences in radii to confirm the HMTs.

\begin{figure}[!thb]
\vspace{-10mm}
	\includegraphics[scale=0.45]{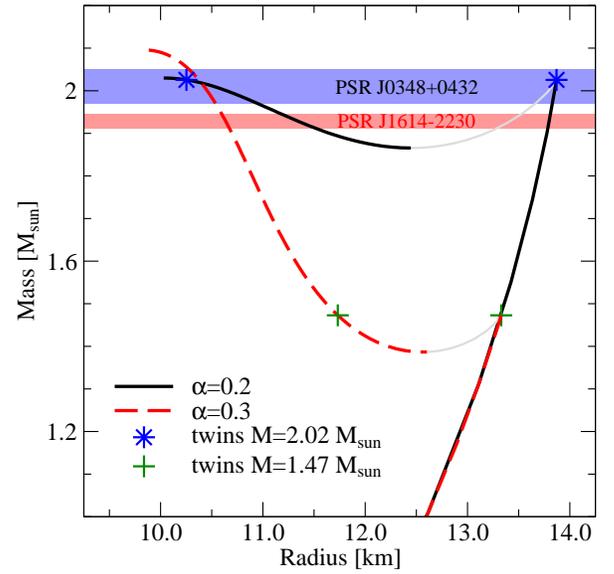}
\vspace{-10mm}
	\caption{(Color online) M-R relations for different values of the available volume $\alpha$, while $a = -1.0 \mathrm{\;MeV\;fm^3}$, $b=2.0 \mathrm{\;MeV\;fm^9}$, and $c=0.036 \mathrm{\;fm^6}$ are kept fixed. The hadronic phase is described by the DD2p40 EoS. The grey lines represent the unstable configurations of hybrid stars. The points on the plot indicate a possible twin solution for each branch.}
	\label{fig:thisone}
\end{figure}

The energy shift at transition density, finely-tuned by the coupling parameter $a$ of the 4-quark interaction, has a subtle effect on the onset of the transition to quark matter. The maximum mass of the purely hadronic neutron star is decreased as the parameter value is increased, see Fig.~\ref{fig:vary_a_TOV}. 

The energy shift at high densities, dominated by the coupling parameter $b$ of the 8-quark interaction, 
has two effects. The repulsive interaction potential provided to the EoS by this term significantly stiffens the quark phase of the matter. The stiffer quark core is capable of supporting much more massive configurations. A consequence of the stiffer quark-matter phase is that the transition to the quark-matter phase occurs much later. This results in a greater purely hadronic neutron star mass at the transition point, see Fig.~\ref{fig:vary_b_TOV}.

\begin{table}[!h]
\label{tab:masses}
\begin{tabular}{|l|c|c|}
\hline
&$\alpha=0.2$ & $\alpha=0.3$ \\
\hline \hline
$M_{\rm max}^{NS}$ [$M_\odot$] & 2.023 & 1.471 \\
$R^{\rm NS}$ [km] &13.87 & 13.34\\
$R^{\rm HS}$ [km] &10.26 & 11.73\\
$\Delta R$ [km] &3.61 & 1.61\\
\hline
$M_{\rm min}^{HS}$ [$M_\odot$] & 1.864 & 1.383 \\
$R^{\rm NS}$ [km] &13.75 & 13.21\\
$R^{\rm HS}$ [km] &12.44 & 12.56\\
$\Delta R$ [km] &1.31 & 0.65\\
\hline
$M_{\rm max}^{HS}$ [$M_\odot$] & 2.030 & 2.097 \\
$R^{\rm HS}$ [km] &10.04 & 9.87\\
\hline
\end{tabular}
\caption{
Maximum ($M_{\rm max}^{NS}$) and minimum ($M_{\rm min}^{HS}$) masses of twin star configurations and the corresponding pair of radii resulting in a radius difference $\Delta R = R^{NS} - R^{HS}$ for different values of the available volume parameter $\alpha=0.2$ (middle column) and  $\alpha=0.3$ (right column). }
\end{table}

In order to produce an accessible (e.g., by mass accretion or spin down of a maximum mass neutron star from the hadronic branch) and stable hybrid star branch, a sufficiently rapid drop of the quark mass within the chiral symmetry restoration transition is necessary, since it leads to a stiffening of the EoS with increasing density. 
In our approach, the dynamical quark mass is determined by the scalar selfenergy (\ref{eqnshifts2}), 
based on a confining interaction with the string tension being medium dependent  due to the available volume correction (\ref{Dv}). 
The parameter $\alpha$ determines the density dependence of chiral symmetry restoration, see 
Fig.~\ref{fig:mass_vs_nb}.
It is evident in Fig.~\ref{fig:vary_avol_TOV} that without the restoration (or even too slow of a restoration) the hybrid configuration is unstable against collapse. Proper consideration of the available volume for quark matter supports the existence of high-mass neutron stars in the range of measured stars \cite{Demorest:2010bx, Antoniadis:2013pzd}. 
Of particular interest are two parametrized equations of state with differing values of the $\alpha$ parameter.
The EoS are analyzed further with a focus on the twin solutions produced within.

\begin{figure}[!htb]
\vspace{-10mm}
	\includegraphics[scale=0.45]{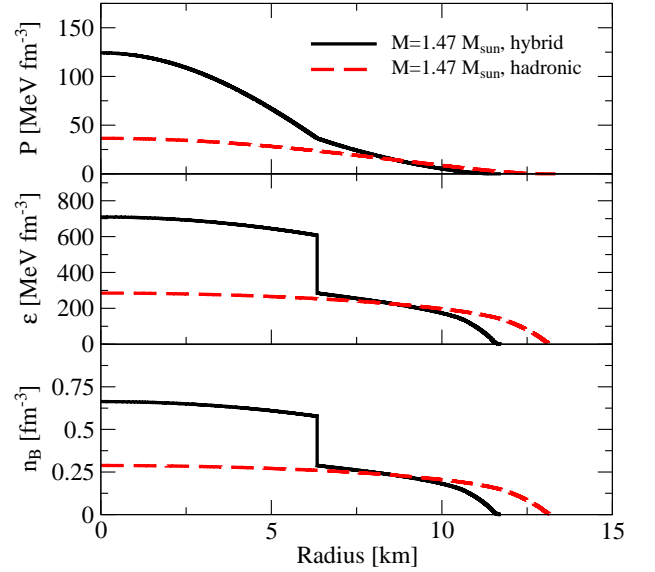}
\vspace{-10mm}
	\caption{(Color online) The radial profiles 
of pressure (upper panel), energy density (middle panel) and baryon density (lower panel) for the twin solutions at $M=1.47M_{\odot}$ for $\alpha=0.2$, indicated in Fig.~\ref{fig:thisone} by plus signs. Dashed lines correspond to the purely hadronic (NS) solution, while the solid lines are for the hybrid star case.}
	\label{fig:profiles145}
\end{figure}

\begin{figure}[!htb]
\vspace{-10mm}
	\includegraphics[scale=0.45]{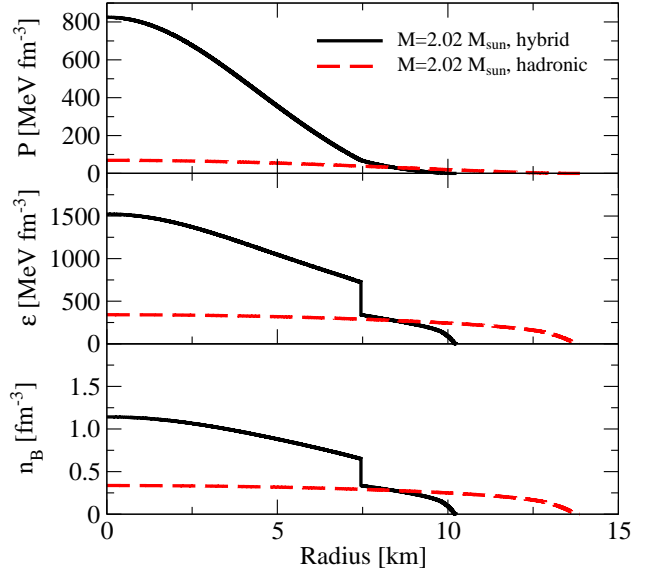}
\vspace{-10mm}
	\caption{(Color online) 
         Same as Fig.~\ref{fig:profiles145}, for the high-mass twin solutions with $M=2.02~M_{\odot}$ 
	obtained for $\alpha=0.3$ and indicated in Fig.~\ref{fig:thisone} by a star sign.}
	\label{fig:profiles202}
\end{figure}

The larger value of $\alpha$ (as seen Fig.~\ref{fig:thisone}) has a profound effect on the critical, transition pressure, where the matter goes over from the hadronic phase to the quark phase. The rate at which the available volume of the confining nonperturbative QCD vacuum is reduced allows for twin solutions to exist at much lower compact star masses, for which the majority of the observational data pertains.
The values $\alpha=0.2$ and $\alpha=0.3$ correspond to two interesting cases for neutron star phenomenology. They correspond to an onset of deconfinement at $M_{\rm max}^{NS}=2.023~M_\odot$ 
and  $M_{\rm max}^{NS}=1.471~M_\odot$, respectively.
Both cases produce third family sequences of hybrid stars (HS), separated from the purely hadronic 
neutron star (NS) branch by a sequence of unstable solutions (indicated by the grey lines in 
Fig.~\ref{fig:thisone}).
While in the former case the hybrid stars (and therefore twin star solutions) are restricted to the high-mass domain $1.864 < M^{HS}[M_\odot]<2.030$ and the radius difference $\Delta R$ of the twins amounts to 
$1.31 < \Delta R [$km$] < 3.61$, in the latter case the NS branch ends already  at 
$M_{\rm max}^{NS}=1.471~M_\odot$ and the hybrid star solutions cover a large band of masses   
$1.383 < M^{HS}[M_\odot]<2.097$, with a smaller radius difference for the twins  
$0.65 < \Delta R [$km$] < 1.61$. 
While in the former case hybrid star solutions are restricted to the high-mass domain, in the latter case 
the purely hadronic neutron star solutions are restricted to masses below $1.471~M_\odot$  and all stars
more massive than $1.383~M_\odot$ can be hybrid ones. 
These results are summarized in Table~\ref{tab:masses}.
Note that the chosen values for $\alpha$ serve for an orientation on the spectrum of solutions that are possible within the present class of high-density quark-matter EoS.

In Figs.~\ref{fig:profiles145} and \ref{fig:profiles202}, we compare the density profiles of pressure, energy density and baryon density for the twin star solutions indicated in Fig.~\ref{fig:thisone} by plusses and by stars,
respectively. While the purely hadronic stars (red dashed lines) have a monotonic profile of these quantities at a rather low level, the hybrid star profiles (black solid lines) exhibit jumps in energy density and baryon density by about a factor two at about one half (for $\alpha=0.2$ and $M=1.47~M_\odot$) and three quarters  (for $\alpha=0.3$ and $M=2.02~M_\odot$).

In particular in the case $\alpha=0.2$ with high-mass twins as those shown in Fig.~\ref{fig:profiles202}, there is a significant compactification when comparing the $NS$ and the $HS$ branches with 
a radius difference up to $\Delta R=3.6$ km that could be detectable with instruments like NICER 
\cite{NICER1}.
Besides different radii and therefore different momenta of inertia, there are qualitative differences expected in, e.g., the cooling behavior of high-mass twins, resulting from their different internal composition and largely different densities in their cores. 

\begin{figure}[!htb]
\vspace{-5mm}
	\includegraphics[scale=0.45]{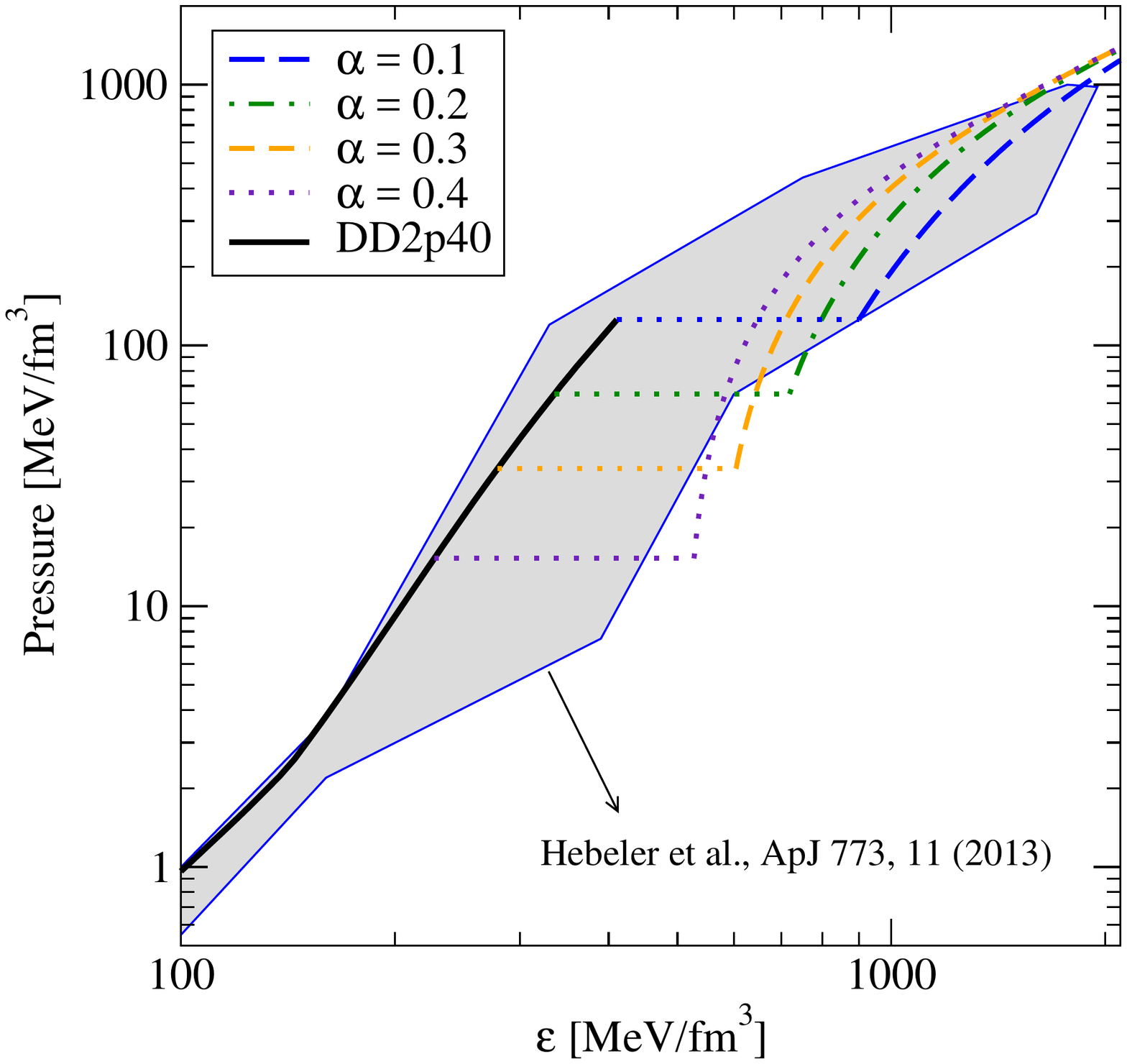}
\vspace{-10mm}
	\caption{(Color online) The EoS shown in Fig.~\ref{fig:eos-edens_vary_avol} compared to the EoS constraint region analyzed with a multi-polytrope ansatz for the high-density EoS by
	Ref.~\cite{Hebeler:2013nza}.}
	\label{fig:Hebeler}
\end{figure}

As a final point of discussion, we compare our hybrid EoS with the variation of the available volume parameter $\alpha$ to a more conservative analysis of the EoS constraint that follows from a multi-polytrope ansatz for the high-density EoS in Fig.~\ref{fig:Hebeler}. We find that the cases discussed
as most favorable hybrid EoS in this work are in fair agreement with the shaded region in the plane
of pressure vs. energy density that was deduced by Hebeler et al. in Ref.~\cite{Hebeler:2013nza}.
The parameters of such a multi-polytrope EoS can be adjusted so that a strong first-order phase transition is described which entails the HMT phenomenon, as it has recently been demonstrated in
Ref.~\cite{Alvarez-Castillo:2017qki}.
In comparison, the parameters in the EoS of the present work can be linked to their microscopic origin.
With such an EoS, one could potentially gain insights on the microphysics of strongly interacting matter
by measuring masses and radii of neutron stars to a sufficient accuracy.

\section{Conclusion}
\label{Conclusion}

We have presented an effective relativistic density functional approach to the quark-nuclear hybrid matter
and applied it to the EoS at zero temperature for the purpose of studying compact stars, in particular the HMT phenomenon.
The mechanism of chiral quark confinement is based on a Cornell-type confining potential.
To this end, we have revived the string-flip model for a practical application within compact stars.
We have introduced an available volume fraction to the quark-matter phase, to mirror the excluded volume in the hadronic matter, and have exhibited the important effect that such a consideration has on the stiffness of the quark matter.
The effective reduction of the string tension in dense matter by the available volume fraction
$\Phi(\bar{n}_v)$ results in an appropriate softening of the EoS in the region of the hadronic to quark phase-transition.
This approach satisfies observational constraints for pulsars taking into consideration a transition to quark matter, while at the same time successfully fulfilling the requirements to obtain a stable third branch of hybrid stars and HMT solutions.
Our approach has been limited to considering the third branch of hybrid stars. 
Recently, Alford and Sedrakian  have considered a synthetic EoS for high-density quark matter which results in a ``fourth family" of hybrid stars, based on a second first order phase transition from two-flavor quark matter to three-flavor color-superconducting (CFL) quark matter, see \cite{Alford:2017qgh}.
The present approach can be extended to consider such questions based on an appropriately chosen density functional. 
It can also be straightforwardly  generalized to finite temperatures so that quark-hadron hybrid equations of state for applications in simulations of supernova explosions and neutron star mergers can be provided.

\acknowledgments
The authors acknowledge fruitful discussions with D. Alvarez-Castillo, D. Ebert, T. Fischer, H. Grigorian,
T. Kl\"ahn, P. M. Lo, M. Marczenko, G. R\"opke, L. Turko and S. Typel.
This work was supported by NCN grant UMO-2014/13/B/ST9/02621.

\end{document}